\documentclass[12pt]{article}
\usepackage{setspace}
\usepackage{amsmath}
\usepackage{graphicx}
\usepackage{endfloat}
\usepackage{amssymb}
\usepackage[default]{jasa_harvard}
\usepackage{JASA_manu}
\usepackage{color}

\newcommand{\thelemma}{{\thesection. \arabic{lemma}}}

\newtheorem{theorem}{{\sc Theorem}}
\newtheorem{lemma}{{\sc Lemma}}
\newtheorem{corollary}{{\sc Corollary}}

\input{tcilatex}

\newcommand{\CO}{{\mathcal{O}}}
\newcommand{\CU}{{\mathcal{U}}}

\newcommand{\E}{{\mbox{\rm E}}}

\def\Co{{\scriptstyle \mathcal{O}}} 
\def\Cu{{\scriptstyle \mathcal{U}}} 
\def\T{{\top}} 

\begin{document}

\title{Statistical Inference for Generalized Additive Partially Linear Model \footnote{This is a post-peer-review, pre-copyedit version of an article published in the Journal of Multivariate Analysis. The final authenticated version is available online at: http://dx.doi.org/10.1016/j.jmva.2017.07.011}}

\author{ Rong \textsc{Liu} \\
Department of Mathematics and Statistics\\
University of Toledo, OH\\
email: \texttt{rong.liu@utoledo.edu} \\
Wolfgang K. \textsc{H\"{a}rdle} \\
Center for Applied Statistics and Economics\\
Humboldt-Universit\"{a}t zu Berlin, Germany \\
and \\
School of Business \\
Singapore Management University, Singapore\\
email: \texttt{haerdle@wiwi.hu-berlin.de} \\
Guoyi \textsc{Zhang} \\
Department of Mathematics and Statistics \\
The University of New Mexico, NM\\
email: \texttt{gzhang@unm.edu} }
\date{}
\maketitle

\newpage

\begin{center}
\textbf{Abstract}
\end{center}

The Generalized Additive Model (GAM) is a powerful tool and has been well
studied. This model class helps to identify additive regression structure.
Via available test procedures one may identify the regression structure even
sharper if some component functions have parametric form. The Generalized
Additive Partially Linear Models (GAPLM) enjoy the simplicity of the GLM and
the flexibility of the GAM because they combine both parametric and
nonparametric components. We use the hybrid spline-backfitted kernel
estimation method, which combines the best features of both spline and
kernel methods for making fast, efficient and reliable estimation under $%
\alpha $\textit{-}mixing condition. In addition, simultaneous confidence
corridors (SCCs) for testing overall trends and empirical likelihood
confidence region for parameters are provided under independent condition.
The asymptotic properties are obtained and simulation results support the
theoretical properties. For the application, we use the GAPLM to improve the
accuracy ratio of the default predictions for $19610$ German companies. The
quantlet for this paper are available on https://github.com.



\newpage

\section{Introduction}

\smallskip The class of generalized additive models (GAMs) provides an
effective semiparametric regression tool for high dimensional data, see [6]. 
{For a response $Y$ and a predictor vector $\mathbf{X}=\left(
X_{1},\ldots,X_{d}\right) ^{\top }$, the pdf of $Y_{i}$ conditional on $%
\mathbf{X}_{i}$ with respect to a fixed $\sigma $-finite measure from
exponential families is%
\begin{equation*}
f\left( Y_{i}\left\vert \mathbf{X}_{i},\phi \right. \right) =\exp \left[
\left\{ Y_{i}m\left( \mathbf{X}_{i}\right) -b\left\{ m\left( \mathbf{X}%
_{i}\right) \right\} \right\} /a\left( \phi \right) +h\left( Y_{i},\phi
\right) \right] .
\end{equation*}%
The function$\ b$ is a given function which relates $m\left( \mathbf{x}%
\right) $ to the conditional variance function $\sigma ^{2}\left( \mathbf{x}%
\right) =\mathrm{var} \left( Y|\mathbf{X=x}\right) $ via the equation $%
\sigma ^{2}\left( \mathbf{x}\right) =a\left( \phi \right) b^{\prime \prime
}\left\{ m\left( \mathbf{x}\right) \right\} $, in which $a\left( \phi
\right) $ is a nuisance parameter that quantifies overdispersion. For the
theoretical development, it is not necessary to assume that the data $%
\left\{ Y_{i},\mathbf{X}_{i}^{\top }\right\} _{i=1}^{n}$ come from such an
exponential family, but only that the conditional variance and conditional
mean are linked by the following equation%
\begin{equation*}
\mathrm{var} \left( Y|\mathbf{X=x}\right) =a\left( \phi \right) b^{\prime
\prime }\left[ \left( b^{\prime }\right) ^{-1}\left\{ \E\left( Y|\mathbf{X=x}%
\right) \right\} \right] .
\end{equation*}%
More specifically, the model is 
\begin{equation}
\E\left( Y|\mathbf{X}\right) =b^{\prime }\left\{ c+\tsum\nolimits_{\alpha
=1}^{d}m_{\alpha }\left( X_{\alpha }\right) \right\} ,  \label{DEF:GAM}
\end{equation}%
with $b^{\prime }$ is the derivative of function $b$. } Model (\ref{DEF:GAM}%
) can for example be used in scoring methods and analyzing default of
companies (Here $Y=1$ denotes default and $b^{\prime }=e^{y}/1+e^{y}$ is the
link function ). Fitting Model (\ref{DEF:GAM}) to such a default data set
leads to $d$ estimated component functions $\hat{m}_{\alpha }\left( \cdot
\right) $ was studied in [11, 25]. Plotting these $\hat{m}_{\alpha }\left(
\cdot \right) $ with simultaneous confidence corridors (SCCs) as developed
by [25], one can check the functional form and therefore obtain simpler
parameterizations of $m_{\alpha }$.

The typical approach is to perform a preliminary (nonparametric) analysis on
the influence of the component functions, and one may improve the model by
introducing parametric components. This will lead to simplification, more
interpretability and higher precision in statistical calibration. With these
thoughts in mind, the GAM model changes to a Generalized Additive Partially
Linear Model (GAPLM): 
\begin{equation}
\E\left( Y|\mathbf{T,X}\right) =b^{\prime }\left\{ m\left( \mathbf{T},%
\mathbf{X}\right) \right\} ,  \label{DEF:GPLM}
\end{equation}%
with $m\left( \mathbf{T},\mathbf{X}\right) =\mathbf{\beta }^{\top }\mathbf{T+%
}\sum_{\alpha =1}^{d_{2}}m_{\alpha }\left( X_{\alpha }\right) $ and%
\begin{equation*}
\mathbf{\beta }=\left( \beta _{0},\beta _{1},\ldots,\beta _{d_{1}}\right)
^{\top }\mathbf{,T}=\left( T_{0},T_{1},\ldots,T_{d_{1}}\right) ^{\top },%
\mathbf{X}=\left( X_{1},\ldots,X_{d_{2}}\right) ^{\top },
\end{equation*}%
where $T_{0}=1$, $T_{k}\mathbf{\in }\mathbb{R}$ for $1\leq k\leq d_{1}$. In
this paper, we have following equation%
\begin{equation*}
\mathrm{var} \left( Y|\mathbf{T=t,X=x}\right) =a\left( \phi \right)
b^{\prime \prime }\left[ \left( b^{\prime }\right) ^{-1}\left\{ \E\left( Y|%
\mathbf{T=t,X=x}\right) \right\} \right] .
\end{equation*}%
We can write (\ref{DEF:GPLM}) in the usual regression form:%
\begin{equation*}
Y_{i}=b^{\prime }\left\{ m\left( \mathbf{T}_{i},\mathbf{X}_{i}\right)
\right\} +\sigma \left( \mathbf{T}_{i},\mathbf{X}_{i}\right) \varepsilon _{i}
\end{equation*}%
with white noise $\varepsilon _{i}$ that satisfies $\E\left( \varepsilon
_{i}|\mathbf{T}_{i},\mathbf{X}_{i}\right) =0,$ $\E\left( \varepsilon
_{i}^{2}|\mathbf{T}_{i},\mathbf{X}_{i}\right) =1$. For identifiability,%
\begin{equation}
\E\left\{ m_{\alpha }\left( X_{\alpha }\right) \right\} =0,1\leq \alpha \leq
d_{2}.  \label{constrain}
\end{equation}%
As in most works on nonparametric smoothing, estimation of the functions $%
\left\{ m_{\alpha }\left( x_{\alpha }\right) \right\} _{\alpha =1}^{d_{2}}$
is conducted on compact sets. Without lose of generality, let the compact
set be $\mathbf{\varkappa }=\left[ 0,1\right] ^{d_{2}}$.

Some estimation methods for Model (\ref{DEF:GPLM}) have been proposed, but
are either computationally expensive or lacking theoretical justification.
The kernel-based backfitting and marginal integration methods e.g., in [3,
9, 24], are computationally expensive. In the meanwhile, more advanced non-
and semiparametric models (without link function) have been studied, such as
partially linear model and varying-coefficient model, see [10, 12, 16, 21,
22]. [21] proposed a nonconcave penalized quasi-likelihood method, with
polynomial spline smoothing for estimation of $m_{\alpha },1\leq \alpha \leq
d_{2}$, and deriving quasi-likelihood based estimators for the linear
parameter $\mathbf{\beta \in }\mathbb{R}^{1+d_{1}}$. To our knowledge, [21]
is a pilot paper since it provides asymptotic normality of the estimators
for the parametric components in GAPLM with independent observations.
However, asymptotic normality for estimations of the nonparametric component
functions $m_{\alpha },1\leq \alpha \leq d_{2}$ and SCCs are still missing.
Recently, [13] studied more complicated Generalized Additive Coefficient
Model by using two-step spline method, but independent and identical
assumptions are required for the asymptotic properties of the estimation and
inference of $m_{\alpha }$, and the asymptotic normality of parameter
estimations is also missing. [5]{\ developed nonparametric analysis of
deviance tools, which can be used to test the significance of the
nonparametric term in generalized partially linear models with univariate
nonparametric component function. [8] provided empirical likelihood based
confidence region for parameter} $\mathbf{\beta }$ and pointwise confidence
interval for nonparametric term in generalized partially linear models.

The spline backfitted kernel (SBK) estimation introduced in [20] combines
the advantages of both kernel and spline methods and the result is balanced
in terms of theory, computation, and interpretation. The basic idea is to
pre-smooth the component functions by spline estimation and then use the
kernel method to improve the accuracy of the estimation on a specific $%
m_{\alpha }$. {In this paper we extend the SBK method to calibrate Model (%
\ref{DEF:GPLM}) with additive nonparametric components, as a result we
obtain oracle efficiency and asymptotic normality of the estimators for both
the parametric and nonparametric components under $\alpha $\textit{-}mixing
condition, which complicates the proof of the theoretical properties. With
stronger i.i.d assumption, we provide empirical likelihood (EL) based
confidence region for parameter} $\mathbf{\beta }$ due to the advantages of
EL such as increase of accuracy of coverage, easy implementation, avoiding
estimating variances and studentising automatically{, see [8]. In addition
we provide SCCs for the nonparametric component functions based on maximal
deviation distribution in [2] so one can test the hypothesis of the shape
for nonparametric terms. }

The paper is organized as follows. In Section 2, we discuss the details of (%
\ref{DEF:GPLM}). In Section 3, the oracle estimator and their asymptotic
properties are introduced. In Section 4, the SBK estimator is introduced and
the asymptotics for both the parametric and nonparametric component
estimations are given. In addition, SCCs for testing overall trends and
entire shapes are considered. In Section 5, we apply the methods to
simulated and real data examples. All technical proofs are given in the
Appendix.

\section{Model assumptions\label{sec:assumptions}}

The space of $\alpha $-centered square integrable functions on $[0,1]$ is
defined as in [18],

\begin{equation*}
\mathcal{H}_{\alpha }^{0}=\left\{ g:\E\left\{ g\left( X_{\alpha }\right)
\right\} =0,\E\left\{ g^{2}\left( X_{\alpha }\right) \right\} <+\infty
\right\} .
\end{equation*}%
Next define the model space $\mathcal{M}$, a collection of functions on $%
\mathbb{R}^{d_{2}}$ as 
\begin{equation*}
\mathcal{M}=\left\{ g\left( \mathbf{x}\right) =\tsum\nolimits_{\alpha
=1}^{d_{2}}g_{\alpha }\left( \mathbf{x}\right) ;g_{\alpha }\in \mathcal{H}%
_{\alpha }^{0}\right\} .
\end{equation*}%
The constraints that $\E\left\{ g_{\alpha }\left( X_{\alpha }\right)
\right\} =0$, $1\leq \alpha \leq d_{2}$ ensure the unique additive
representation of $m_{\alpha }$ as expressed in (\ref{constrain}). Denote
the empirical expectation by $\E_{n}$, then $\E_{n}\varphi
=\sum_{i=1}^{n}\varphi \left( \mathbf{X}_{i}\right) /n$. For functions $%
g_{1},g_{2}\in \mathcal{M}$, the theoretical and empirical inner products
are defined respectively as $\left\langle g_{1},g_{2}\right\rangle =\E%
\left\{ g_{1}\left( \mathbf{X}\right) g_{2}\left( \mathbf{X}\right) \right\} 
$, $\left\langle g_{1},g_{2}\right\rangle _{n}=\E_{n}\left\{ g_{1}\left( 
\mathbf{X}\right) g_{2}\left( \mathbf{X}\right) \right\} $. The
corresponding induced norms are $\left\Vert g_{1}\right\Vert _{2}^{2}=\E %
g_{1}^{2}\left( \mathbf{X}\right) $, $\left\Vert g_{1}\right\Vert _{2,n}^{2}=%
\E_{n}g_{1}^{2}\left( \mathbf{X}\right) $. More generally, we define $%
\left\Vert g\right\Vert _{r}^{r}=\E\left\vert g\left( \mathbf{X}\right)
\right\vert ^{r}.$

In the paper, for any compact interval $[a,b]$, we denote the space of $p$%
-th order smooth functions as $C^{\left( p\right) }[a,b]=\left\{ g|g^{\left(
p\right) }\in C\left[ a,b\right] \right\} $, and the class of Lipschitz
continuous functions for constant $C>0$ as $\limfunc{Lip}\left( \left[ a,b%
\right] ,C\right) =\left\{ g|\left\vert g\left( x\right) -g\left( x^{\prime
}\right) \right\vert \leq C\left\vert x-x^{\prime }\right\vert \text{, }%
\forall x,x^{\prime }\in \left[ a,b\right] \right\} $. For any vector $%
\mathbf{x}=\left( x_{1},x_{2},\cdots ,x_{d}\right) ^{\top}$, we denote the
supremum and $p$ norms as $\left\vert \mathbf{x}\right\vert =\mathrm{max}%
_{1\leq \alpha \leq d}\left\vert x_{\alpha }\right\vert $ and $\left\Vert 
\mathbf{x}\right\Vert _{p}=\left( \sum_{\alpha =1}^{d}x_{\alpha }^{p}\right)
^{1/p}$. In particular, we use $\left\Vert \mathbf{x}\right\Vert $ to denote
the Euclidean norm, i.e., $p=2$. We need the following assumptions:

\begin{enumerate}
\item[(A1)] \textit{The additive component functions }$m_{\alpha }\in
C^{\left( 1\right) }\left[ 0,1\right] ,1\leq \alpha \leq d_{2}$\textit{\
with }$m_{1}\in C^{\left( 2\right) }\left[ 0,1\right] $, $m_{\alpha
}^{\prime }\in \limfunc{Lip}\left( \left[ 0,1\right] ,C_{m}\right) $, $2\leq
\alpha \leq d_{2}$\textit{\ for some constant }$C_{m}>0$\textit{.}

\item[(A2)] \textit{The inverse link function }$b^{\prime }$\textit{\
satisfies: }$b^{\prime }\in C^{2}\left( \mathbb{R}\right) ,b^{\prime \prime
}\left( \theta \right) >0,\theta \in \mathbb{R}$\textit{\ and }$C_{b}>%
\mathrm{max}_{\theta \in \Theta }b^{\prime \prime }\left( \theta \right)
\geq \mathrm{min}_{\theta \in \Theta }b^{\prime \prime }\left( \theta
\right) >c_{b}$\textit{\ for constants }$C_{b}>c_{b}>0$\textit{. }

\item[(A3)] \textit{The conditional variance function} $\sigma ^{2}\left( 
\mathbf{x}\right) $ \textit{is measurable and bounded. The errors }$\left\{
\varepsilon _{i}\right\} _{i=1}^{n}$\textit{\ satisfy }$\E\left( \varepsilon
_{i}|\mathcal{F}_{i}\right) =0,$\textit{\ }$\E\left( \left\vert \varepsilon
_{i}\right\vert ^{2+\eta }\right) \leq C_{\eta }$\textit{\ for some }$\eta
\in \left( 1/2,+\infty \right) $\textit{\ with the sequence of }$\sigma $%
\textit{-fields: } \newline
$\mathcal{F}_{i}=\sigma \left\{ \left( \mathbf{X}_{j}\right) ,j\leq
i;\varepsilon _{j},j\leq i-1\right\} $\textit{\ for }$i=1,\ldots ,n$.

\item[(A4)] \textit{The density function }$f\left( \mathbf{x}\right) $%
\textit{\ of }$\left( X_{1},\ldots,X_{d_{2}}\right) $\textit{\ is continuous
and} 
\begin{equation*}
0<c_{f}\leq \mathrm{inf}_{\mathbf{x}\in \mathbf{\chi }}f\left( \mathbf{x}%
\right) \leq \mathrm{sup}_{\mathbf{x}\in \mathbf{\varkappa }}f\left( \mathbf{%
x}\right) \leq C_{f}<\infty .
\end{equation*}%
\textit{The marginal densities }$f_{\alpha }\left( x_{\alpha }\right) $%
\textit{\ of }$X_{\alpha }$\textit{\ have continuous derivatives on }$\left[
0,1\right] $\textit{\ as well as the uniform upper bound }$C_{f}$\textit{\
and lower bound }$c_{f}$\textit{.}

\item[(A5)] \textit{Constants }$K_{0},\lambda _{0}\in \left( 0,+\infty
\right) $\textit{\ exist such that }$\alpha \left( n\right) \leq
K_{0}e^{-\lambda _{0}n}$ \textit{holds for all }$n$\textit{, with the }$%
\alpha $\textit{-mixing coefficients for }$\left\{ \mathbf{Z}_{i}=\left( 
\mathbf{T}_{i}^{\top},\mathbf{X}_{i}^{\top},\varepsilon _{i}\right) ^{\T%
}\right\} _{i=1}^{n}$\ \textit{\ defined as }%
\begin{equation*}
\alpha \left( k\right) =\mathrm{sup}_{B\in \sigma \left\{ \mathbf{Z}%
_{s},s\leq t\right\} ,C\in \sigma \left\{ \mathbf{Z}_{s},s\geq t+k\right\}
}\left\vert \limfunc{P}\left( B\cap C\right) -\limfunc{P}\left( B\right) 
\limfunc{P}\left( C\right) \right\vert ,k\geq 1.
\end{equation*}

\item[(A5')] $\left\{ \mathbf{Z}_{i}=\left( \mathbf{T}_{i}^{\top},\mathbf{X}%
_{i}^{\top},\varepsilon _{i}\right) ^{\top}\right\} _{i=1}^{n}$\textit{\ are
independent and identically distributed.}

\item[(A6)] There exist constants $0<c_{\delta }<C_{\delta }<\infty $ and $%
0<c_{\mathbf{Q}}<C_{\mathbf{Q}}<\infty $ such that $c_{\delta }\leq \E\left(
\left\vert T_{k}\right\vert ^{2+\delta }|\mathbf{X=x}\right) \leq C_{\delta
} $ for some $\delta >0,$ and $c_{\mathbf{Q}}I_{d_{1}\times d_{1}}\leq \E%
\left( \mathbf{TT}^{\top}|\mathbf{X=x}\right) \leq C_{\mathbf{Q}%
}I_{d_{1}\times d_{1}}$ .
\end{enumerate}

Assumptions (A1), (A2) and (A4) are standard in the GAM literature, see [19,
23], while Assumptions (A3) and (A5) are the same for weakly dependent data
as in [11, 20] and Assumption (A6) is the same with (C5) in [21]. When
categorical predictors presents, we can create dummy variables in $\mathbf{T}%
_{i}$ and Assumption (A6) is still satisfied.

\section{Oracle estimators\label{sec:smoother}}

The aim of our analysis is to provide precise estimators for the component
functions $m_{\alpha }\left( \cdot \right) $ and parameters $\mathbf{\beta }$%
. Without loss of generality, we may focus on $m_{1}\left( \cdot \right) $.
If all the unknown $\mathbf{\beta }$ and other $\left\{ m_{\alpha }\left(
x_{\alpha }\right) \right\} _{\alpha =2}^{d_{2}}$ were known, we are in a
comfortable situation since the multidimensional modelling problem has
reduced to one dimension. As in [17], define for each $x_{1}\in \left[ h,1-h%
\right] $, $a\in A$ a local quasi log-likelihood function 
\begin{equation*}
\tilde{\ell}_{m_{1}}\left( a,x_{1}\right) =n^{-1}\tsum\nolimits_{i=1}^{n}%
\left[ Y_{i}\left\{ a+m\left( \mathbf{T}_{i},\mathbf{X}_{i\_1}\right)
\right\} -b\left\{ a+m\left( \mathbf{T}_{i},\mathbf{X}_{i\_1}\right)
\right\} \right] K_{h}\left( X_{i1}-x_{1}\right)
\end{equation*}%
with $m\left( \mathbf{T}_{i},\mathbf{X}_{i\_1}\right) =\mathbf{\beta }^{\top
}\mathbf{T}_{i}+\tsum\nolimits_{\alpha =2}^{d_{2}}m_{\alpha }\left( \mathbf{X%
}_{i\alpha }\right) $ and $K_{h}\left( u\right) =K\left( u/h\right) /h$ a
kernel function $K$ with bandwidth $h$ that satisfy

\begin{enumerate}
\item[(A7)] \textit{The kernel function} $K\left( \cdot \right) $\textit{\ }$%
\in C^{1}[-1,1]$ \textit{is a symmetric pdf.} \textit{The bandwidth }$%
h=h_{n} $\textit{\ satisfies }$h=\Co\left\{ n^{-1/5}(\log n)^{-1/5}\right\}
, $ $h^{-1}=\CO\left\{ n^{1/5}\left( \ln n\right) ^{\delta }\right\} $ 
\textit{for some constant }$\delta >1/5$\textit{.}
\end{enumerate}

{Since all the {$\mathbf{\beta }$ and $\left\{ m_{\alpha }\left( x_{\alpha
}\right) \right\} _{\alpha =2}^{d_{2}}$ are known as obtained from oracle, }%
one can obtain the so-called oracle estimator } 
\begin{equation}
\tilde{m}_{K,1}\left( x_{1}\right) =\func{argmax}_{a\in A}\tilde{\ell}%
_{m_{1}}\left( a,x_{1}\right) .  \label{DEF:mtilde}
\end{equation}%
Denote $\left\Vert K\right\Vert _{2}^{2}=\int K^{2}\left( u\right) du$, $\mu
_{2}\left( K\right) =\int K\left( u\right) u^{2}du$ and the scale function $%
D_{1}\left( x_{1}\right) $ and bias function $\limfunc{bias}%
\nolimits_{1}\left( x_{1}\right) $%
\begin{equation}
D_{1}\left( x_{1}\right) =f_{1}\left( x_{1}\right) \E\left\{ b^{\prime
\prime }\left\{ m\left( \mathbf{T},\mathbf{X}\right) \right\}
|X_{1}=x_{1}\right\} ,  \label{DEF:D1x1}
\end{equation}%
\begin{eqnarray}
\limfunc{bias}{}_{1}\left( x_{1}\right) &=&\mu _{2}\left( K\right) \left[
m_{1}^{\prime \prime }\left( x_{1}\right) f_{1}\left( x_{1}\right) \E\left[
b^{\prime \prime }\left\{ m\left( \mathbf{T},\mathbf{X}\right) \right\}
|X_{1}=x_{1}\right] \right.  \notag \\
&&+m_{1}^{\prime }\left( x_{1}\right) \frac{\partial }{\partial x_{1}}%
\left\{ f_{1}\left( x_{1}\right) \E\left[ b^{\prime \prime }\left\{ m\left( 
\mathbf{T},\mathbf{X}\right) \right\} |X_{1}=x_{1}\right] \right\}  \notag \\
&&\left. -\left\{ m_{1}^{\prime }\left( x_{1}\right) \right\}
^{2}f_{1}\left( x_{1}\right) \E\left[ b^{\prime \prime \prime }\left\{
m\left( \mathbf{T},\mathbf{X}\right) \right\} |X_{1}=x_{1}\right] \right] .
\label{DEF:b1x1}
\end{eqnarray}

\begin{lemma}
\label{THM:oracleasydist}Under Assumptions (A1)-(A7), for any $x_{1}\in %
\left[ h,1-h\right] $, as $n\rightarrow \infty $, the oracle kernel
estimator $\tilde{m}_{\limfunc{K},1}\left( x_{1}\right) $ given in (\ref%
{DEF:mtilde}) satisfies%
\begin{equation*}
\mathrm{sup}_{x_{1}\in \left[ h,1-h\right] }\left\vert \tilde{m}_{\limfunc{K}%
,1}\left( x_{1}\right) -m_{1}\left( x_{1}\right) \right\vert =\CO%
_{a.s.}\left( \log n/\sqrt{nh}\right) ,
\end{equation*}%
\begin{equation*}
\sqrt{nh}\left\{ \tilde{m}_{\limfunc{K},1}\left( x_{1}\right) -m_{1}\left(
x_{1}\right) -\limfunc{bias}\nolimits_{1}\left( x_{1}\right)
h^{2}/D_{1}\left( x_{1}\right) \right\} \overset{\tciLaplace }{\rightarrow }%
N\left( 0,D_{1}\left( x_{1}\right) ^{-1}v_{1}^{2}\left( x_{1}\right)
D_{1}\left( x_{1}\right) ^{-1}\right) ,
\end{equation*}%
with%
\begin{equation*}
v_{1}^{2}\left( x_{1}\right) =f_{1}\left( x_{1}\right) \E\left\{ \sigma
^{2}\left( \mathbf{T},\mathbf{X}\right) |X_{1}=x_{1}\right\} \left\Vert
K\right\Vert _{2}^{2}.
\end{equation*}
\end{lemma}

Lemma 1 is given in [11]. The above oracle idea applies to the parametric
part as well. Define the log-likelihood function%
\begin{equation}
\tilde{\ell}_{\mathbf{\beta }}\left( \mathbf{a}\right)
=n^{-1}\sum\nolimits_{i=1}^{n}\left[ Y_{i}\left\{ \mathbf{a}^{\top }\mathbf{T%
}_{i}+m\left( \mathbf{X}_{i}\right) \right\} -b\left\{ \mathbf{a}^{\top }%
\mathbf{T}_{i}+m\left( \mathbf{X}_{i}\right) \right\} \right] ,
\label{DEF:ltilde}
\end{equation}%
where $m\left( \mathbf{X}_{i}\right) =\sum_{\alpha =1}^{d_{2}}m_{\alpha
}\left( X_{i\alpha }\right) $. The infeasible estimator of $\mathbf{\beta }$
is $\mathbf{\tilde{\beta}}=\func{argmax}_{\mathbf{a}\in \mathbb{R}^{1+d_{1}}}%
\tilde{\ell}_{\mathbf{\beta }}\left( \mathbf{a}\right) $. Clearly, $\nabla 
\tilde{\ell}_{\mathbf{\beta }}\left( \mathbf{\beta }\right) =\mathbf{0}$. To
maximize (\ref{DEF:ltilde})\bigskip , we have%
\begin{equation*}
n^{-1}\tsum\nolimits_{i=1}^{n}\left[ Y_{i}\mathbf{T}_{i}-b^{\prime }\left\{ 
\mathbf{a}^{\top }\mathbf{T}_{i}+m\left( \mathbf{X}_{i}\right) \right\} 
\mathbf{T}_{i}\right] =\mathbf{0,}
\end{equation*}%
then the empirical likelihood ratio is%
\begin{equation*}
\tilde{R}\left( \mathbf{a}\right) =\mathrm{max} \left\{ \Pi
_{i=1}^{n}np_{i}|\Sigma _{i=1}^{n}p_{i}Z_{i}\left( \mathbf{a}\right) =%
\mathbf{0},p_{i}\geq 0,\Sigma _{i=1}^{n}p_{i}=1\right\}
\end{equation*}%
where $\ Z_{i}\left( \mathbf{a}\right) =\left[ Y_{i}-b^{\prime }\left\{ 
\mathbf{a}^{\top }\mathbf{T}_{i}+m\left( \mathbf{X}_{i}\right) \right\} %
\right] \mathbf{T}_{i}$.

\begin{theorem}
\label{THM:betatilde-beta}(i) Under Assumptions (A1)-(A6), as $n\rightarrow
\infty ,$%
\begin{equation*}
\left\vert \mathbf{\tilde{\beta}}-\mathbf{\beta }-\left[ \E b^{\prime \prime
}\left\{ m\left( \mathbf{T},\mathbf{X}\right) \right\} \mathbf{TT}^{\top}%
\right] ^{-1}n^{-1}\tsum\nolimits_{i=1}^{n}\sigma \left( \mathbf{T}_{i},%
\mathbf{X}_{i}\right) \varepsilon _{i}\mathbf{T}_{i}\right\vert =\CO%
_{a.s.}\left( n^{-1}\left( \log n\right) ^{2}\right) ,
\end{equation*}%
\begin{equation*}
\sqrt{n}\left( \mathbf{\tilde{\beta}}-\mathbf{\beta }\right) \overset{%
\tciLaplace }{\rightarrow }N\left( \mathbf{0},a\left( \phi \right) \left[ \E %
b^{\prime \prime }\left\{ m\left( \mathbf{T},\mathbf{X}\right) \right\} 
\mathbf{TT}^{\top}\right] ^{-1}\right) .
\end{equation*}%
(ii) Under Assumptions (A1)-(A4), (A5') and (A6), 
\begin{equation*}
-2\log \tilde{R}\left( \mathbf{\beta }\right) \overset{\tciLaplace }{%
\rightarrow }\chi _{d_{1}}^{2}.
\end{equation*}
\end{theorem}

Although the oracle estimators $\mathbf{\tilde{\beta}}$ and $\tilde{m}%
_{K,1}\left( x_{1}\right) $ enjoy the desirable theoretical properties in
Theorem \ref{THM:betatilde-beta} and Lemma \ref{THM:oracleasydist}, they are
not a feasible statistic as its computation is based on the knowledge of
unavailable component functions $\left\{ m_{\alpha }\left( x_{\alpha
}\right) \right\} _{\alpha =2}^{d_{2}}$.

\section{Spline-backfitted kernel estimators\label{sec:estimator}}

In practice, the rest components $\left\{ m_{\alpha }\left( x_{\alpha
}\right) \right\} _{\alpha =2}^{d_{2}}${\ are of course unknown and need to
be approximated. We obtain the spline-backfitted kernel estimators by using\
estimations of $\left\{ m_{\alpha }\left( x_{\alpha }\right) \right\}
_{\alpha =2}^{d_{2}}$ and the unknown $\mathbf{\beta }$ by splines and we
employ them to estimate $m_{1}\left( x_{1}\right) $ as in (\ref{DEF:mtilde}).%
} First, we introduce the linear spline basis as in [10]. Let $0=\xi
_{0}<\xi _{1}<\cdots <\xi _{N}<\xi _{N+1}=1$ denote a sequence of equally
spaced points, called interior knots, on $\left[ 0,1\right] $. Denote by $%
H=\left( N+1\right) ^{-1}$ the width of each subinterval $\left[ \xi
_{J},\xi _{J+1}\right] ,0\leq J\leq N$ and denote the degenerate knots $\xi
_{-1}=0,\xi _{N+2}=1$. We need the following assumption:

\begin{enumerate}
\item[(A8)] \textit{The number of interior knots} $N\thicksim n^{1/4}\log n,$%
\textit{\ i.e.,} $c_{N}n^{1/4}\log n\leq N\leq C_{N}n^{1/4}\log n$\textit{\
for some constants }$c_{N}$,$C_{N}$\textit{\ }$>0$.
\end{enumerate}

\noindent Following [11], for $J=0,\ldots ,N+1$, define the linear B spline
basis:%
\begin{equation*}
b_{J}\left( x\right) =\left( 1-\left\vert x-\xi _{J}\right\vert /H\right)
_{+}=\left\{ 
\begin{array}{c}
\left( N+1\right) x-J+1 \\ 
J+1-\left( N+1\right) x \\ 
0%
\end{array}%
\begin{array}{c}
, \\ 
, \\ 
,%
\end{array}%
\begin{array}{c}
\xi _{J-1}\leq x\leq \xi _{J} \\ 
\xi _{J}\leq x\leq \xi _{J+1} \\ 
\text{otherwise}%
\end{array}%
\right. ,
\end{equation*}%
the space of $\alpha $-empirically centered linear spline functions on $%
[0,1]:$%
\begin{equation*}
G_{n,\alpha }^{0}=\left\{ g_{\alpha }:g_{\alpha }\left( x_{\alpha }\right)
=\tsum\nolimits_{J=0}^{N+1}\lambda _{J}b_{J}\left( x_{\alpha }\right) ,\E%
_{n}\left\{ g_{\alpha }\left( X_{\alpha }\right) \right\} =0\right\} ,1\leq
\alpha \leq d_{2},
\end{equation*}%
and the space of additive spline functions on $\mathbf{\chi }$: 
\begin{equation*}
G_{n}^{0}=\left\{ g\left( \mathbf{x}\right) =\tsum\nolimits_{\alpha
=1}^{d_{2}}g_{\alpha }\left( x_{\alpha }\right) ;g_{\alpha }\in G_{n,\alpha
}^{0}\right\} .
\end{equation*}%
Define the log-likelihood function 
\begin{equation}
\hat{L}\left( \mathbf{\beta ,}g\right) =n^{-1}\sum\nolimits_{i=1}^{n}\left[
Y_{i}\left\{ \mathbf{\beta }^{\top }\mathbf{T}_{i}\mathbf{+}g\left( \mathbf{X%
}_{i}\right) \right\} -b\left\{ \mathbf{\beta }^{\top }\mathbf{T}%
_{i}+g\left( \mathbf{X}_{i}\right) \right\} \right] ,g\in G_{n}^{0},
\label{DEF:splinelikelihood}
\end{equation}%
which according to Lemma 14 of [19], has a unique maximizer with probability
approaching $1$. The multivariate function $m\left( \mathbf{x}\right) $ is
then estimated by the additive spline function $\hat{m}\left( \mathbf{x}%
\right) $ with%
\begin{equation*}
\hat{m}\left( \mathbf{t},\mathbf{x}\right) =\mathbf{\hat{\beta}}^{\top }%
\mathbf{t+}\hat{m}\left( \mathbf{x}\right) =\func{argmax}_{g\in G_{n}^{0}}%
\hat{L}\left( \mathbf{\beta ,}g\right) .
\end{equation*}%
Since $\hat{m}\left( \mathbf{x}\right) \in G_{n}^{0}$, one can write $\hat{m}%
\left( \mathbf{x}\right) =\sum_{\alpha =1}^{d_{2}}\hat{m}_{\alpha }\left(
x_{\alpha }\right) $ for $\hat{m}_{\alpha }\left( x_{\alpha }\right) \in
G_{n,\alpha }^{0}$. Next define the log-likelihood function%
\begin{equation}
\hat{\ell}_{m_{1}}\left( a,x_{1}\right) =n^{-1}\tsum\nolimits_{i=1}^{n}\left[
Y_{i}\left\{ a+\hat{m}\left( \mathbf{T}_{i},\mathbf{X}_{i\_1}\right)
\right\} -b\left\{ a+\hat{m}\left( \mathbf{T}_{i},\mathbf{X}_{i\_1}\right)
\right\} \right] K_{h}\left( X_{i1}-x_{1}\right)  \label{DEF:lhat}
\end{equation}%
where $\hat{m}\left( \mathbf{T}_{i},\mathbf{X}_{i\_1}\right) =\mathbf{\hat{%
\beta}}^{\top }\mathbf{T}_{i}+\sum_{\alpha =2}^{d_{2}}\hat{m}_{\alpha
}\left( X_{i\alpha }\right) $. Define the SBK estimator as:%
\begin{equation}
\hat{m}_{\func{SBK},1}\left( x_{1}\right) =\func{argmax}_{a\in A}\hat{\ell}%
_{m_{1}}\left( a,x_{1}\right) .  \label{DEF:mstarhat}
\end{equation}

\begin{theorem}
\label{THM:mhat-mtilde}Under Assumptions (A1)-(A8), as $n\rightarrow \infty $%
, $\hat{m}_{\func{SBK},1}\left( x_{1}\right) $ is oracally efficient,%
\begin{equation*}
\mathrm{sup}_{x_{1}\in \lbrack 0,1]}\left\vert \hat{m}_{\func{SBK},1}\left(
x_{1}\right) -\tilde{m}_{K,1}\left( x_{1}\right) \right\vert =\CO%
_{a.s.}\left( n^{-1/2}\log n\right) .
\end{equation*}
\end{theorem}

The following corollary is a consequence of Lemma \ref{THM:oracleasydist}
and Theorem \ref{THM:mhat-mtilde}.

\begin{corollary}
\label{COR:SBKproperties}Under Assumptions (A1)-(A8), as $n\rightarrow
\infty ,$ the SBK estimator $\hat{m}_{\func{SBK},1}\left( x_{1}\right) $
given in (\ref{DEF:mstarhat}) satisfies%
\begin{equation*}
\mathrm{sup}_{x_{1}\in \left[ h,1-h\right] }\left\vert \hat{m}_{\func{SBK}%
,1}\left( x_{1}\right) -m_{1}\left( x_{1}\right) \right\vert =\CO%
_{a.s.}\left( \log n/\sqrt{nh}\right)
\end{equation*}%
and for any $x_{1}\in \left[ h,1-h\right] $, with $\limfunc{bias}%
\nolimits_{1}\left( x_{1}\right) $ as in (\ref{DEF:b1x1}) and $D_{1}\left(
x_{1}\right) $ in (\ref{DEF:D1x1})%
\begin{equation*}
\sqrt{nh}\left\{ \hat{m}_{\func{SBK},1}\left( x_{1}\right) -m_{1}\left(
x_{1}\right) -\limfunc{bias}\nolimits_{1}\left( x_{1}\right)
h^{2}/D_{1}\left( x_{1}\right) \right\} \overset{\tciLaplace }{\rightarrow }%
N\left( 0,D_{1}\left( x_{1}\right) ^{-1}v_{1}^{2}\left( x_{1}\right)
D_{1}\left( x_{1}\right) ^{-1}\right) .
\end{equation*}
\end{corollary}

Denote $a_{h}=\sqrt{-2\mathop{\rm{log}}h},C\left( K\right) =\left\Vert
K^{\prime }\right\Vert _{2}^{2}\left\Vert K\right\Vert _{2}^{-2}$ and for
any $\alpha \in \left( 0,1\right) $, the quantile 
\begin{equation*}
Q_{h}(\alpha )=a_{h}+a_{h}^{-1}\left[ \mathop{\rm{log}}\left\{ \sqrt{C\left(
K\right) }/\left( 2\pi \right) \right\} -\mathop{\rm{log}}\left\{ -%
\mathop{\rm{log}}\sqrt{1-\alpha }\right\} \right] .
\end{equation*}%
Also with $D_{1}\left( x_{1}\right) $ and $v_{1}^{2}\left( x_{1}\right) $
given in (\ref{DEF:D1x1}), define%
\begin{equation*}
\sigma _{n}\left( x_{1}\right) =n^{-1/2}h^{-1/2}v_{1}\left( x_{1}\right)
D_{1}^{-1}\left( x_{1}\right) .
\end{equation*}

\begin{theorem}
\label{THM:bands}Under Assumptions (A1)-(A4), (A5'), (A6)-(A8), as $%
n\rightarrow \infty ,$%
\begin{equation*}
\lim_{n\rightarrow \infty }\Pr \left\{ \mathrm{sup}_{x_{1}\in \lbrack
h,1-h]}\left\vert \hat{m}_{\func{SBK},1}\left( x_{1}\right) -m_{1}\left(
x_{1}\right) \right\vert /\sigma _{n}\left( x_{1}\right) \leq Q_{h}\left(
\alpha \right) \right\} =1-\alpha .
\end{equation*}%
A $100\left( 1-\alpha \right) \%$ simultaneous confidence band for $%
m_{1}\left( x_{1}\right) $ is 
\begin{equation*}
\hat{m}_{\func{SBK},1}\left( x_{1}\right) \pm \sigma _{n}\left( x_{1}\right)
Q_{h}\left( \alpha \right) .
\end{equation*}
\end{theorem}

In fact, $\mathbf{\hat{\beta}}$ obtained by maximizing (\ref%
{DEF:splinelikelihood}) is equivalent to $\mathbf{\hat{\beta}}_{\func{SBK}}=%
\func{argmax}_{\mathbf{a}\in \mathbb{R}^{1+d_{1}}}\hat{\ell}_{\mathbf{\beta }%
}\left( \mathbf{a}\right) $ with 
\begin{equation*}
\hat{\ell}_{\mathbf{\beta }}\left( \mathbf{a}\right)
=n^{-1}\tsum\nolimits_{i=1}^{n}\left[ Y_{i}\left\{ \mathbf{a}^{\top}\mathbf{T%
}_{i}+\hat{m}\left( \mathbf{X}_{i}\right) \right\} -b\left\{ \mathbf{a}%
^{\top}\mathbf{T}_{i}+\hat{m}\left( \mathbf{X}_{i}\right) \right\} \right]
\end{equation*}%
in which $\hat{m}\left( \mathbf{X}_{i}\right) =\sum_{\alpha =1}^{d_{2}}\hat{m%
}_{\alpha }\left( X_{i\alpha }\right) $. The empirical likelihood ratio is%
\begin{equation*}
\hat{R}\left( \mathbf{a}\right) =\mathrm{max} \left\{ \Pi
_{i=1}^{n}np_{i}|\Sigma _{i=1}^{n}p_{i}\hat{Z}_{i}\left( \mathbf{a}\right) =%
\mathbf{0},p_{i}\geq 0,\Sigma _{i=1}^{n}p_{i}=1\right\}
\end{equation*}%
where $\ \hat{Z}_{i}\left( \mathbf{a}\right) =\left[ Y_{i}-b^{\prime
}\left\{ \mathbf{a}^{\top}\mathbf{T}_{i}+\hat{m}\left( \mathbf{X}_{i}\right)
\right\} \right] \mathbf{T}_{i}$. Similar to Theorem \ref{THM:mhat-mtilde},
the main result shows that the difference between $\mathbf{\hat{\beta}}$ and
its infeasible counterpart $\mathbf{\tilde{\beta}}$ is asymptotically
negligible.

\begin{theorem}
\label{THM:betahat-beta}(i) Under Assumptions (A1)-(A6) and (A8), as $%
n\rightarrow \infty $, $\mathbf{\hat{\beta}}$ is oracally efficient, i.e., $%
\sqrt{n}\left( \hat{\beta}_{k}-\tilde{\beta}_{k}\right) \overset{p}{%
\rightarrow }0$ for $0\leq k\leq d_{1}$ and hence%
\begin{equation*}
\sqrt{n}\left( \mathbf{\hat{\beta}}-\mathbf{\beta }\right) \overset{%
\tciLaplace }{\rightarrow }N\left( \mathbf{0},a\left( \phi \right) \left[ \E %
b^{\prime \prime }\left\{ m\left( \mathbf{T},\mathbf{X}\right) \right\} 
\mathbf{TT}^{\top}\right] ^{-1}\right) .
\end{equation*}%
(ii) Under Assumptions (A1)-(A4), (A5'), (A6) and (A8), as $n\rightarrow
\infty ,$%
\begin{equation*}
\mathrm{sup} \left\vert -2\log \hat{R}\left( \mathbf{\beta }\right) +2\log 
\tilde{R}\left( \mathbf{\beta }\right) \right\vert =\Co_{p}\left( 1\right) ,
\end{equation*}%
so%
\begin{equation*}
-2\log \hat{R}\left( \mathbf{\beta }\right) \overset{\tciLaplace }{%
\rightarrow }\chi _{d_{1}}^{2}.
\end{equation*}
\end{theorem}

As a reviewer pointing out, an obvious advantage of GAPLM over GAM is the
capability of including categorical predictors. Since $m_{\alpha }$ is not a
function of $\mathbf{T}$ in GAPLM, so we can simply create dummy variables
to represent the categorical effects and use spline estimation. [14]
proposed spline estimation combined with categorical kernel functions to
handle the case when function $m_{\alpha }$ depends on categorical
predictors.

\section{Examples\label{sec:examples}}

We have applied the SBK procedure to both simulated (Example 1) and real
(Example 2) data and implemented our algorithms with the following
rule-of-thumb number of interior knots 
\begin{equation*}
N=N_{n}=\mathrm{min} \left( \left\lfloor n^{1/4}\log n\right\rfloor
+1,\left\lfloor n/4d-1/d\right\rfloor -1\right)
\end{equation*}%
which satisfies (A8), i.e., $N=N_{n}\thicksim n^{1/4}\log n$, and ensures
that the number of parameters in the linear least squares problem is less
than $n/4$, i.e., $1+d\left( N+1\right) \leq n/4$. The bandwidth of $%
h_{\alpha }$ is computed as [11] in the asymptotically optimal way.

\subsection{\textbf{Example 1}}

The data are generated from the model 
\begin{equation*}
\Pr (Y=1|\mathbf{T=t,X=x})=b^{\prime }\left\{ \mathbf{\beta }^{\top}\mathbf{%
T+}\sum\nolimits_{\alpha =1}^{d_{2}}m_{\alpha }\left( X_{\alpha }\right)
\right\} ,b^{\prime }\left( x\right) =\frac{e^{x}}{1+e^{x}}
\end{equation*}%
with $d_{1}=2,d_{2}=5,\mathbf{\beta }=\left( \beta _{0},\beta _{1},\beta
_{2}\right) ^\top =\left( 1,1,1,\right) ^{\top},m_{1}\left( x\right)
=m_{2}\left( x\right) =m_{3}\left( x\right) =\sin \left( 2\pi x\right) $, $%
m_{4}\left( x\right) =\Phi \left( 6x-3\right) -0.5$ and $m_{5}\left(
x\right) =x^{2}-1/3$, where $\Phi $ is the standard normal cdf. The
predictors are generated by transforming the following vector autoregression
(VAR) equation for $0\leq r_{1},r_{2}<1,1\leq i\leq n$, $\mathbf{Z}_{0}=%
\mathbf{0}$ 
\begin{eqnarray*}
\mathbf{Z}_{i} &=&r_{1}\mathbf{Z}_{i-1}+\mathbf{\varepsilon }_{i},\mathbf{%
\varepsilon }_{i}\sim \mathcal{N} \left( 0,\Sigma \right) ,\Sigma =\left(
1-r_{2}\right) \mathbf{I}_{d\times d}+r_{2}\mathbf{1}_{d}\mathbf{1}_{d}^{\T%
},d=d_{1}+d_{2},1\leq i\leq n, \\
\mathbf{T}_{i} &=&\left( 1,Z_{i1},\ldots,Z_{id_{1}},\right)
^{\top},X_{i\alpha }=\Phi \left( \sqrt{1-r_{1}^{2}}Z_{i\alpha }\right)
,1\leq i\leq n,1+d_{1}\leq \alpha \leq d_{1}+d_{2},
\end{eqnarray*}%
with stationary $\mathbf{Z}_{i}=\left( Z_{i1},\ldots,Z_{id}\right)
^{\top}\sim N\left\{ 0,\left( 1-r_{1}^{2}\right) ^{-1}\Sigma \right\} ,$ $%
\mathbf{1}_{d}=\left( 1,\ldots,1\right) ^{\top}$ and $\mathbf{I}_{d\times d}$
is the $d\times d$ identity matrix. The $X$ is transformed from $Z$ to
satisfy Assumption (A4). In this study, we selected four scenarios: $r_{1}=0$%
, $r_{2}=0;r_{1}=0.5,$ $r_{2}=0;r_{1}=0$, $r_{2}=0.5;r_{1}=0.5,r_{2}=0.5$.
The parameter $r_{1}$ controls the dependence between observations and $%
r_{2} $ controls the correlation between variables. {In the selected
scenarios, $r_{1}=0$ indicates independent observations and $r_{1}=0.5$ $%
\alpha $-mixing observations, $r_{2}=0$ indicates independent variables and $%
r_{2}=0.5$ correlated variables within each observation. }Define the
empirical relative efficiency of $\hat{\beta}_{1}$ with respect to $\tilde{%
\beta}_{1}$ as $\func{EFF}_{r}\left( \hat{\beta}_{1}\right) =\left\{ \text{%
MSE}\left( \tilde{\beta}_{1}\right) /\text{MSE}\left( \hat{\beta}_{1}\right)
\right\} ^{1/2}.${\ }

\begin{figure}[tbh]
\begin{center}
\begin{picture}(-15,21)
 \put(-250,-360){
 \includegraphics{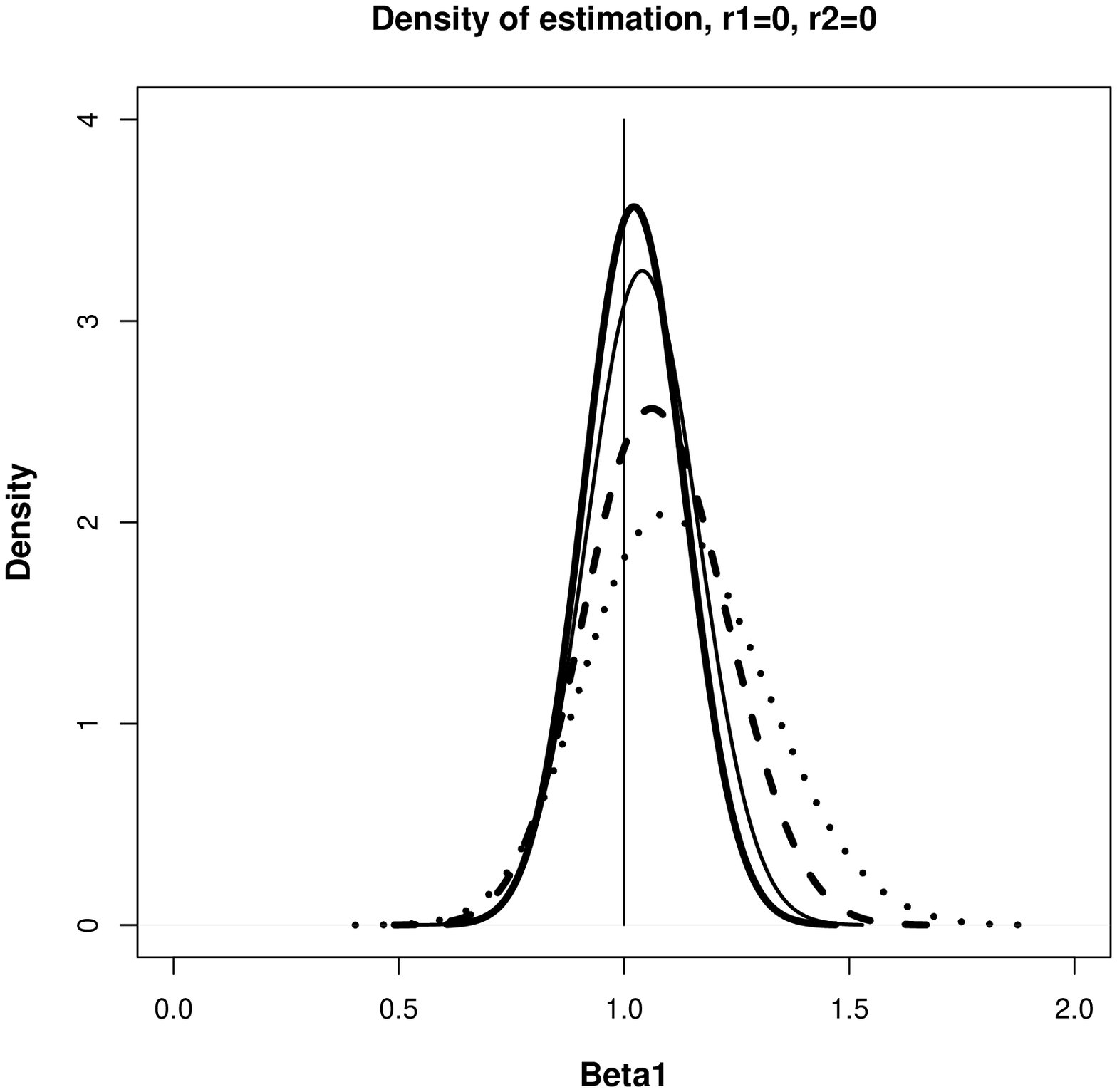}}
 \put(-130,-370){(a)}
 \put(-10,-360){
 \includegraphics{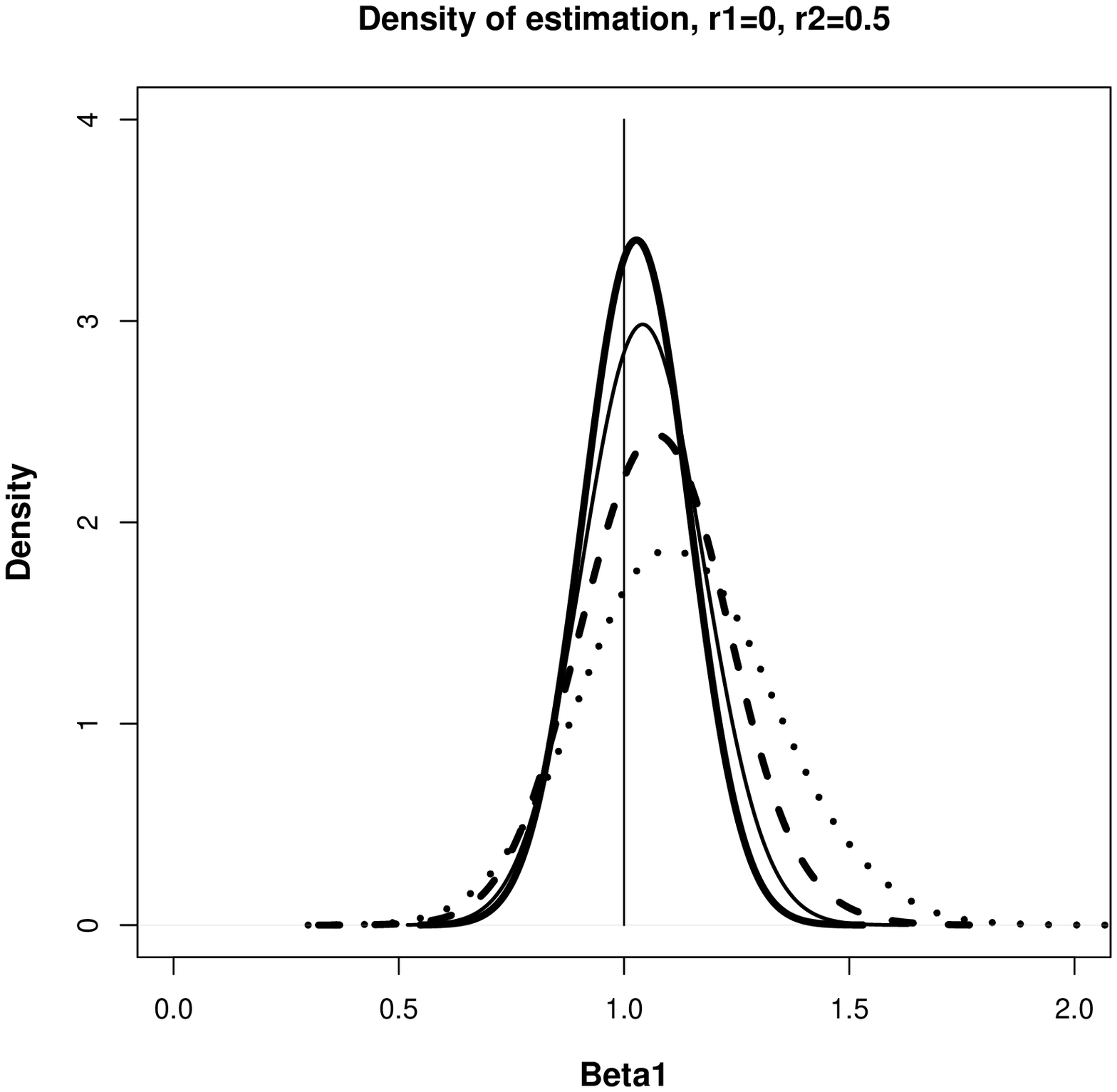}}
 \put(110,-370){(b)}
 \put(-250,-620){
 \includegraphics{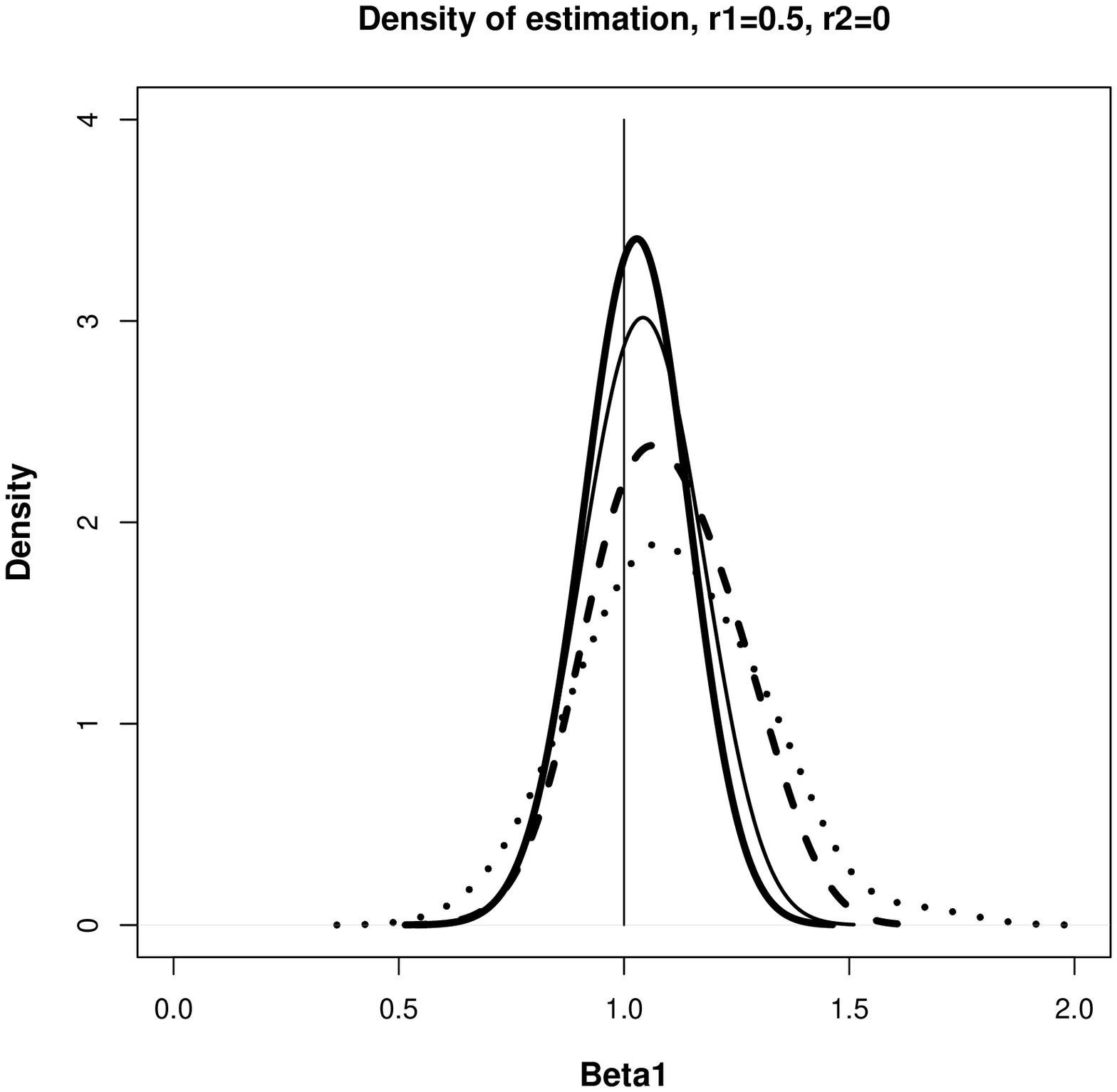}}
 \put(-130,-630){(c)}
 \put(-10,-620){
 \includegraphics{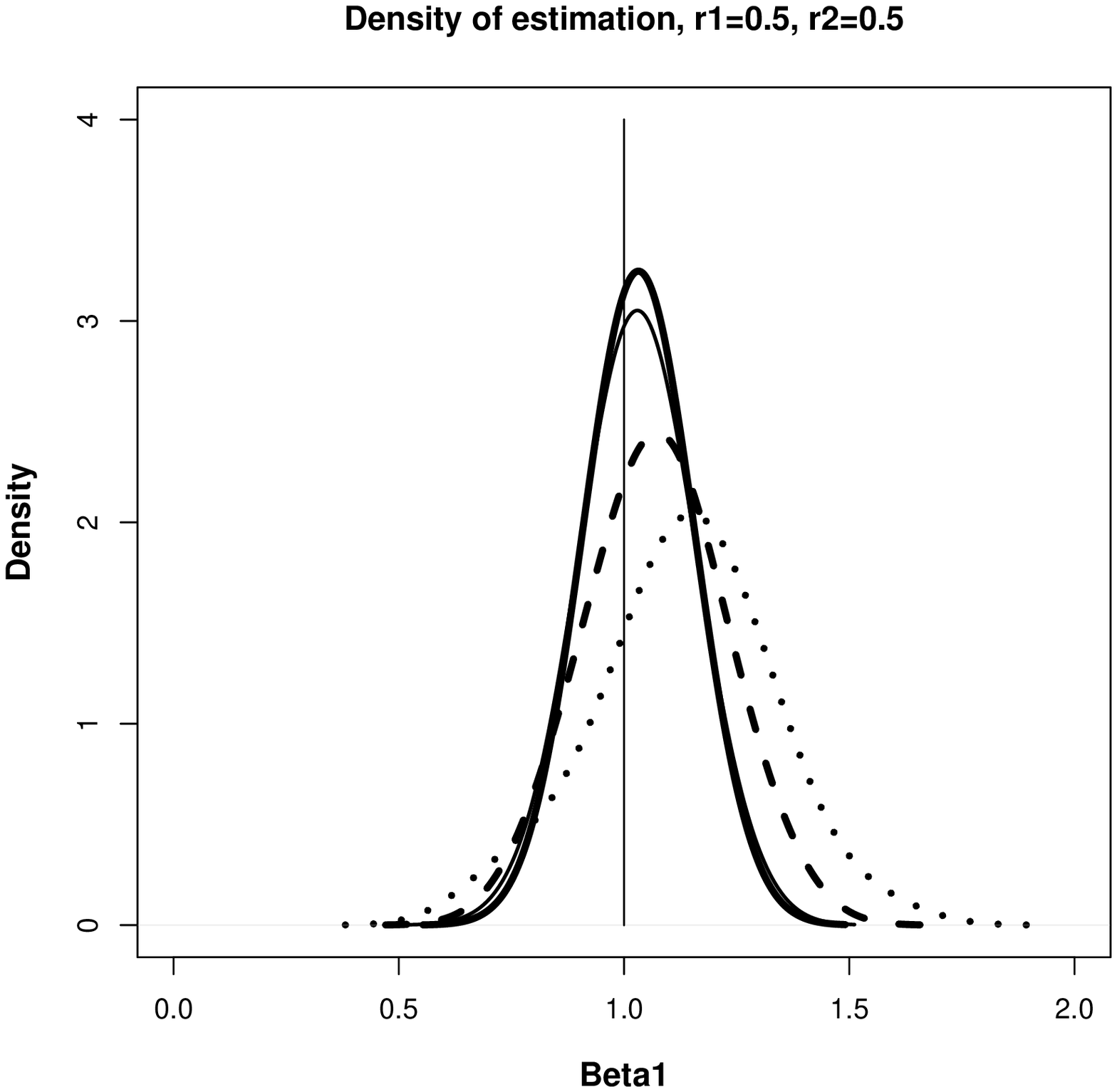}}
 \put(110,-630){(d)}
\end{picture}
\end{center}
\caption{Plots of densities for $\hat{\protect\beta}_{1}$ with $n=500$ -
dotted line, $n=1000$ - dashed line, $n=2000$ - thin solid line, $n=4000$ -
thick solid line for (a) $r_{1}=0,r_{2}=0$, (b) $r_{1}=0,r_{2}=0.5$, (c) $%
r_{1}=0.5,r_{2}=0$, (d) $r_{1}=0.5,r_{2}=0.5$ from $1000$ replications.}
\label{Figuresimu1}
\end{figure}

Table \ref{Tablesimu1} shows the mean of bias, variances, MSEs and EFFs of $%
\hat{\beta}_{1}$ for $R=1000$ with sample sizes $\ n=500,1000,2000,4000$.
The results show that the estimator works as the asymptotic theory
indicates, see Theorem \ref{THM:betahat-beta} (i). 
\begin{table}[th]
\begin{center}
\begin{tabular}{cccccc}
\hline\hline
$r$ & $n$ & $10\times \overline{\text{BIAS}}$ & $100\times \overline{\text{%
VARIANCE}}$ & $100\times \overline{\text{MSE}}$ & $\overline{\func{EFF}}%
\left( \hat{\beta}_{1}\right) $ \\ \hline
\multicolumn{1}{l}{$%
\begin{array}{c}
r_{1}=0 \\ 
r_{2}=0%
\end{array}%
$} & $%
\begin{array}{c}
500 \\ 
1000 \\ 
2000 \\ 
4000%
\end{array}%
$ & $%
\begin{array}{c}
1.509 \\ 
0.727 \\ 
0.408 \\ 
0.240%
\end{array}%
$ & $%
\begin{array}{c}
2.018 \\ 
1.197 \\ 
0.626 \\ 
0.282%
\end{array}%
$ & $%
\begin{array}{c}
4.298 \\ 
1.726 \\ 
0.793 \\ 
0.339%
\end{array}%
$ & $%
\begin{array}{c}
0.8436 \\ 
0.8749 \\ 
0.9189 \\ 
0.9534%
\end{array}%
$ \\ \hline
\multicolumn{1}{l}{$%
\begin{array}{c}
r_{1}=0.5 \\ 
\multicolumn{1}{l}{r_{2}=0}%
\end{array}%
$} & $%
\begin{array}{c}
500 \\ 
1000 \\ 
2000 \\ 
4000%
\end{array}%
$ & $%
\begin{array}{c}
1.473 \\ 
0.834 \\ 
0.476 \\ 
0.260%
\end{array}%
$ & $%
\begin{array}{c}
3.136 \\ 
1.287 \\ 
0.674 \\ 
0.202%
\end{array}%
$ & $%
\begin{array}{c}
5.306 \\ 
1.983 \\ 
0.901 \\ 
0.270%
\end{array}%
$ & $%
\begin{array}{c}
0.8392 \\ 
0.8873 \\ 
0.9294 \\ 
0.9665%
\end{array}%
$ \\ \hline
\multicolumn{1}{l}{$%
\begin{array}{l}
r_{1}=0 \\ 
\multicolumn{1}{c}{r_{2}=0.5}%
\end{array}%
$} & $%
\begin{array}{c}
500 \\ 
1000 \\ 
2000 \\ 
4000%
\end{array}%
$ & $%
\begin{array}{c}
1.327 \\ 
0.699 \\ 
0.665 \\ 
0.390%
\end{array}%
$ & $%
\begin{array}{c}
3.880 \\ 
1.851 \\ 
0.739 \\ 
0.290%
\end{array}%
$ & $%
\begin{array}{c}
5.642 \\ 
2.339 \\ 
1.182 \\ 
0.442%
\end{array}%
$ & $%
\begin{array}{c}
0.8475 \\ 
0.8856 \\ 
0.9353 \\ 
0.9479%
\end{array}%
$ \\ \hline
\multicolumn{1}{l}{$%
\begin{array}{c}
r_{1}=0.5 \\ 
r_{2}=0.5%
\end{array}%
$} & $%
\begin{array}{c}
500 \\ 
1000 \\ 
2000 \\ 
4000%
\end{array}%
$ & $%
\begin{array}{c}
1.635 \\ 
0.901 \\ 
0.529 \\ 
0.209%
\end{array}%
$ & $%
\begin{array}{c}
4.230 \\ 
1.190 \\ 
0.806 \\ 
0.366%
\end{array}%
$ & $%
\begin{array}{c}
6.903 \\ 
2.002 \\ 
1.086 \\ 
0.410%
\end{array}%
$ & $%
\begin{array}{c}
0.8203 \\ 
0.8758 \\ 
0.9304 \\ 
0.9483%
\end{array}%
$ \\ \hline
\end{tabular}%
\end{center}
\caption{The mean of $10\times $Bias, $100\times $Variances, $100\times $%
MSEs and EFFs of$\ \hat{\protect\beta}_{1}$ from $1000$ replications. }
\label{Tablesimu1}
\end{table}
Figure \ref{Figuresimu1} shows the kernel densities of $\hat{\beta}_{1}$s
for $n=500,1000,2000,4000$ from $1000$ replications, again the theoretical
properties are supported. 
Table \ref{Tablesimu5} shows the simulation results of the empirical
likelihood confidence interval for $\beta $ with $n=500,1000,2000,4000$ and $%
r_{1}=0,${$r_{2}=0$} from $1000$ replications. The mean and standard
deviation of $-2\log \hat{R}\left( \mathbf{\beta }\right) +2\log \tilde{R}%
\left( \mathbf{\beta }\right) $ (DIFF) support the oracle efficiency in
Theorem \ref{THM:betahat-beta} (ii). The performance of empirical likelihood
confidence interval are compared with the wald-type one and it is clear that
they have similar performance but empirical likelihood confidence interval
has better coverage ratio and shorter average length. 
\begin{table}[th]
\begin{center}
\begin{tabular}{cccccc}
\hline\hline
&  & $n=500$ & $n=1000$ & $n=2000$ & $n=4000$ \\ \hline
Coverage Ratio & $%
\begin{array}{c}
\text{EL} \\ 
\text{Wald}%
\end{array}%
$ & $%
\begin{array}{c}
0.923 \\ 
0.918%
\end{array}%
$ & $%
\begin{array}{c}
0.941 \\ 
0.934%
\end{array}%
$ & $%
\begin{array}{c}
0.946 \\ 
0.944%
\end{array}%
$ & $%
\begin{array}{c}
0.951 \\ 
0.948%
\end{array}%
$ \\ \hline
Average Length & $%
\begin{array}{c}
\text{EL} \\ 
\text{Wald}%
\end{array}%
$ & $%
\begin{array}{c}
1.2675 \\ 
1.4073%
\end{array}%
$ & $%
\begin{array}{c}
0.9474 \\ 
1.0447%
\end{array}%
$ & $%
\begin{array}{c}
0.7105 \\ 
0.7480%
\end{array}%
$ & $%
\begin{array}{c}
0.5339 \\ 
0.5625%
\end{array}%
$ \\ \hline
$\text{DIFF}$ & $%
\begin{array}{c}
\text{MEAN} \\ 
\text{SD}%
\end{array}%
$ & $%
\begin{array}{c}
0.1213 \\ 
0.5199%
\end{array}%
$ & $%
\begin{array}{c}
0.1023 \\ 
0.4703%
\end{array}%
$ & $%
\begin{array}{c}
0.0981 \\ 
0.3667%
\end{array}%
$ & $%
\begin{array}{c}
0.0726 \\ 
0.3242%
\end{array}%
$ \\ \hline
\end{tabular}%
\end{center}
\caption{Coverage ratios and average length of the empirical likelihood
confidence interval (EL) and Wald-type confidence interval for $\protect%
\beta _{1}$ for $n=500,1000,2000,4000$ with $r_{1}=0$ from $1000$
replications. DIFF$=-2\log \hat{R}\left( \mathbf{\protect\beta }\right)
+2\log \tilde{R}\left( \mathbf{\protect\beta }\right) $ is the difference
between $-2\log \hat{R}\left( \mathbf{\protect\beta }\right) $ and $-2\log 
\tilde{R}\left( \mathbf{\protect\beta }\right) $.}
\label{Tablesimu5}
\end{table}
Next for $\alpha =1,\ldots ,5$, let $X_{\alpha ,\mathrm{min}}^{i}$, $%
X_{\alpha ,\mathrm{max}}^{i}$ denote the smallest and largest observations
of the variable $X_{\alpha }$ in the $i$ -th replication. The component
functions $\left\{ m_{\alpha }\right\} _{\alpha =1}^{5}$ are estimated on
equally spaced points $\left\{ x_{t}\right\} _{t=0}^{100}$ with $%
0=x_{0}<\ldots <x_{100}=1$ and the estimator of $m_{\alpha }$ in the $r$-th
sample\ as $\hat{m}_{\func{SBK},\alpha ,r}$. The (mean) average squared
error (ASE and MASE) are: 
\begin{eqnarray*}
\limfunc{ASE}(\hat{m}_{\func{SBK},\alpha ,r})
&=&101^{-1}\tsum\nolimits_{t=0}^{100}\left\{ \hat{m}_{\func{SBK},\alpha
,r}(x_{t})-m_{\alpha }(x_{t})\right\} ^{2}, \\
\limfunc{MASE}(\hat{m}_{\func{SBK},\alpha })
&=&R^{-1}\tsum\nolimits_{r=1}^{R}\limfunc{ASE}(\hat{m}_{\func{SBK},\alpha
,r}).
\end{eqnarray*}%
In order to examine the efficiency of $\hat{m}_{\func{SBK},\alpha }$
relative to the oracle estimator\ $\tilde{m}_{K,\alpha }\left( x_{\alpha
}\right) $, both are computed using the same data-driven bandwidth $\hat{h}%
_{\alpha ,\limfunc{opt}},$ described in Section 5 of [11]. Define the
empirical relative efficiency of $\hat{m}_{\func{SBK},\alpha }$ with respect
to $\tilde{m}_{K,\alpha }$ as%
\begin{equation*}
\func{EFF}_{r}\left( \hat{m}_{\func{SBK},\alpha }\right) =\left[ \frac{%
\sum\nolimits_{t=0}^{100}\left\{ \tilde{m}_{K,\alpha }\left( x_{t}\right)
-m_{\alpha }(x_{t})\right\} ^{2}}{\sum\nolimits_{t=0}^{100}\left\{ \hat{m}_{%
\func{SBK},\alpha ,r}(x_{t})-m_{\alpha }(x_{t})\right\} ^{2}}\right] ^{1/2}.
\end{equation*}%
EFF measures the relative efficiency of the SBK estimator to the oracle
estimator. For increasing sample size, it should increase to 1 by {Theorem %
\ref{THM:mhat-mtilde}.} Table \ref{Tablesimu2} shows the MASEs of $\tilde{m}%
_{K,1}$, $\hat{m}_{\func{SBK},1}$and the average of EFFs from $1000$
replications for $n=500$, $1000$, $2000$, $4000$. It is clear that the MASEs
of both SBK estimator and the oracle estimator decrease when sample sizes
increase, and the SBK estimator performs as well asymptotically as the
oracle estimator, see Theorem \ref{THM:mhat-mtilde}. 
\begin{table}[th]
\begin{center}
\begin{tabular}{ccccc}
\hline\hline
$r$ & $n$ & $100\times \limfunc{MASE}\left( \tilde{m}_{\limfunc{K},\alpha
}\right) $ & $100\times \limfunc{MASE}\left( \hat{m}_{\func{SBK},\alpha
}\right) $ & $\overline{\func{EFF}}\left( \hat{m}_{\func{SBK},1}\right) $ \\ 
\hline
\multicolumn{1}{l}{$%
\begin{array}{c}
r_{1}=0 \\ 
r_{2}=0%
\end{array}%
$} & $%
\begin{array}{c}
500 \\ 
1000 \\ 
2000 \\ 
4000%
\end{array}%
$ & $%
\begin{array}{c}
4.482 \\ 
2.418 \\ 
1.582 \\ 
1.212%
\end{array}%
$ & $%
\begin{array}{c}
4.603 \\ 
2.503 \\ 
1.613 \\ 
1.247%
\end{array}%
$ & $%
\begin{array}{c}
0.9501 \\ 
0.9809 \\ 
0.9854 \\ 
0.9923%
\end{array}%
$ \\ \hline
\multicolumn{1}{l}{$%
\begin{array}{c}
r_{1}=0.5 \\ 
\multicolumn{1}{l}{r_{2}=0}%
\end{array}%
$} & $%
\begin{array}{c}
500 \\ 
1000 \\ 
2000 \\ 
4000%
\end{array}%
$ & $%
\begin{array}{c}
4.060 \\ 
2.592 \\ 
1.746 \\ 
1.194%
\end{array}%
$ & $%
\begin{array}{c}
4.322 \\ 
2.649 \\ 
1.714 \\ 
1.218%
\end{array}%
$ & $%
\begin{array}{c}
0.9445 \\ 
0.9767 \\ 
0.9832 \\ 
0.9936%
\end{array}%
$ \\ \hline
\multicolumn{1}{l}{$%
\begin{array}{l}
r_{1}=0 \\ 
\multicolumn{1}{c}{r_{2}=0.5}%
\end{array}%
$} & $%
\begin{array}{c}
500 \\ 
1000 \\ 
2000 \\ 
4000%
\end{array}%
$ & $%
\begin{array}{c}
4.845 \\ 
2.935 \\ 
1.951 \\ 
1.515%
\end{array}%
$ & $%
\begin{array}{c}
6.348 \\ 
3.559 \\ 
2.177 \\ 
1.648%
\end{array}%
$ & $%
\begin{array}{c}
0.8827 \\ 
0.8755 \\ 
0.9494 \\ 
0.9795%
\end{array}%
$ \\ \hline
\multicolumn{1}{l}{$%
\begin{array}{c}
r_{1}=0.5 \\ 
r_{2}=0.5%
\end{array}%
$} & $%
\begin{array}{c}
500 \\ 
1000 \\ 
2000 \\ 
4000%
\end{array}%
$ & $%
\begin{array}{c}
5.656 \\ 
2.804 \\ 
1.886 \\ 
1.525%
\end{array}%
$ & $%
\begin{array}{c}
7.114 \\ 
3.570 \\ 
2.089 \\ 
1.634%
\end{array}%
$ & $%
\begin{array}{c}
0.8722 \\ 
0.8951 \\ 
0.9478 \\ 
0.9744%
\end{array}%
$ \\ \hline
\end{tabular}%
\end{center}
\caption{The $100\times $MASEs of $\tilde{m}_{\limfunc{K},1}$, $\hat{m}_{%
\func{SBK},1}$and $\overline{\func{EFF}}$s for $n=500$, $1000$, $2000$, $4000
$ from $1000$ replications.}
\label{Tablesimu2}
\end{table}

\begin{figure}[tbh]
\begin{center}
\begin{picture}(-15,21)
 \put(-250,-360){
 \includegraphics{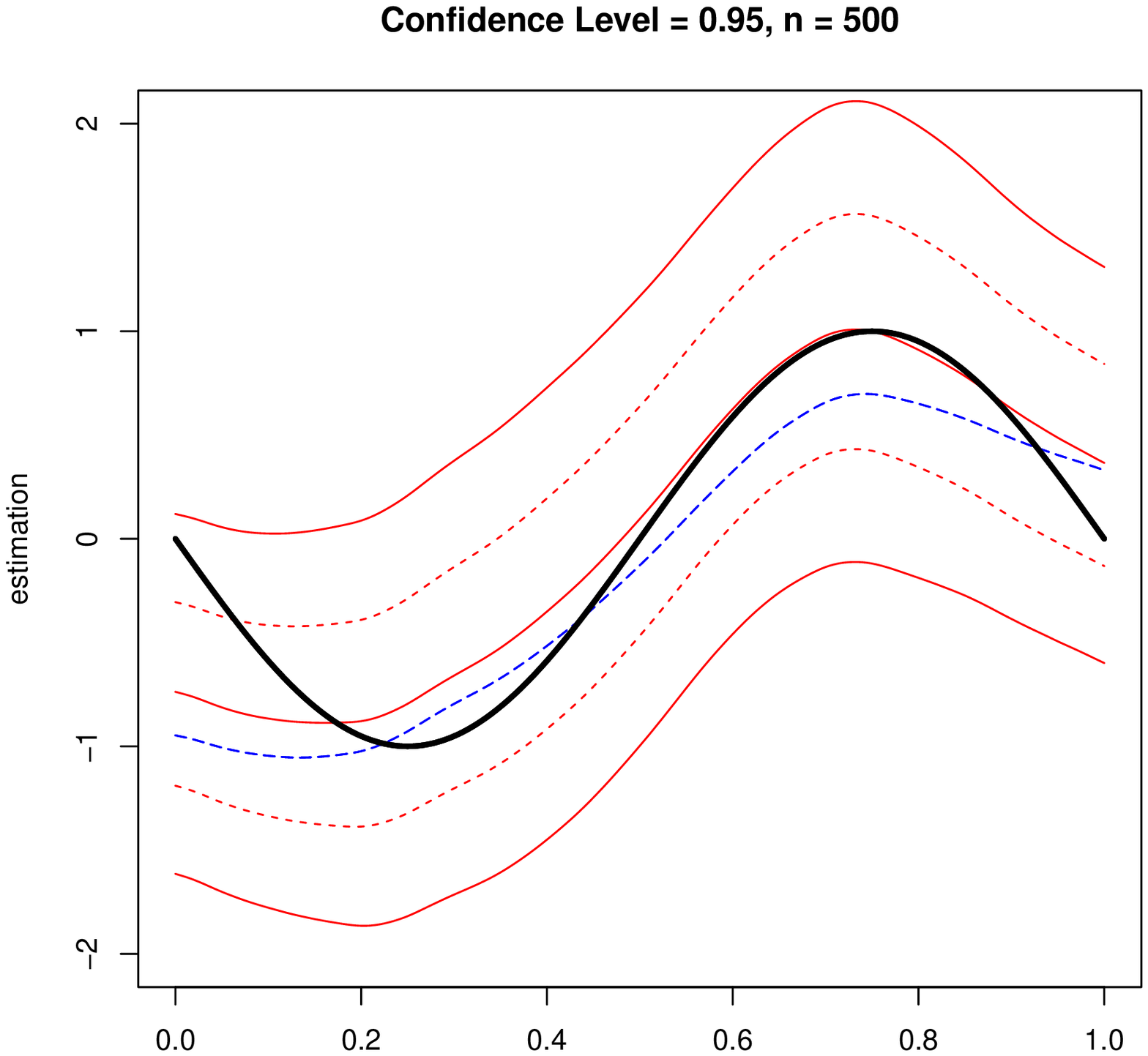}}
 \put(-130,-370){(a)}
 \put(-10,-360){
 \includegraphics{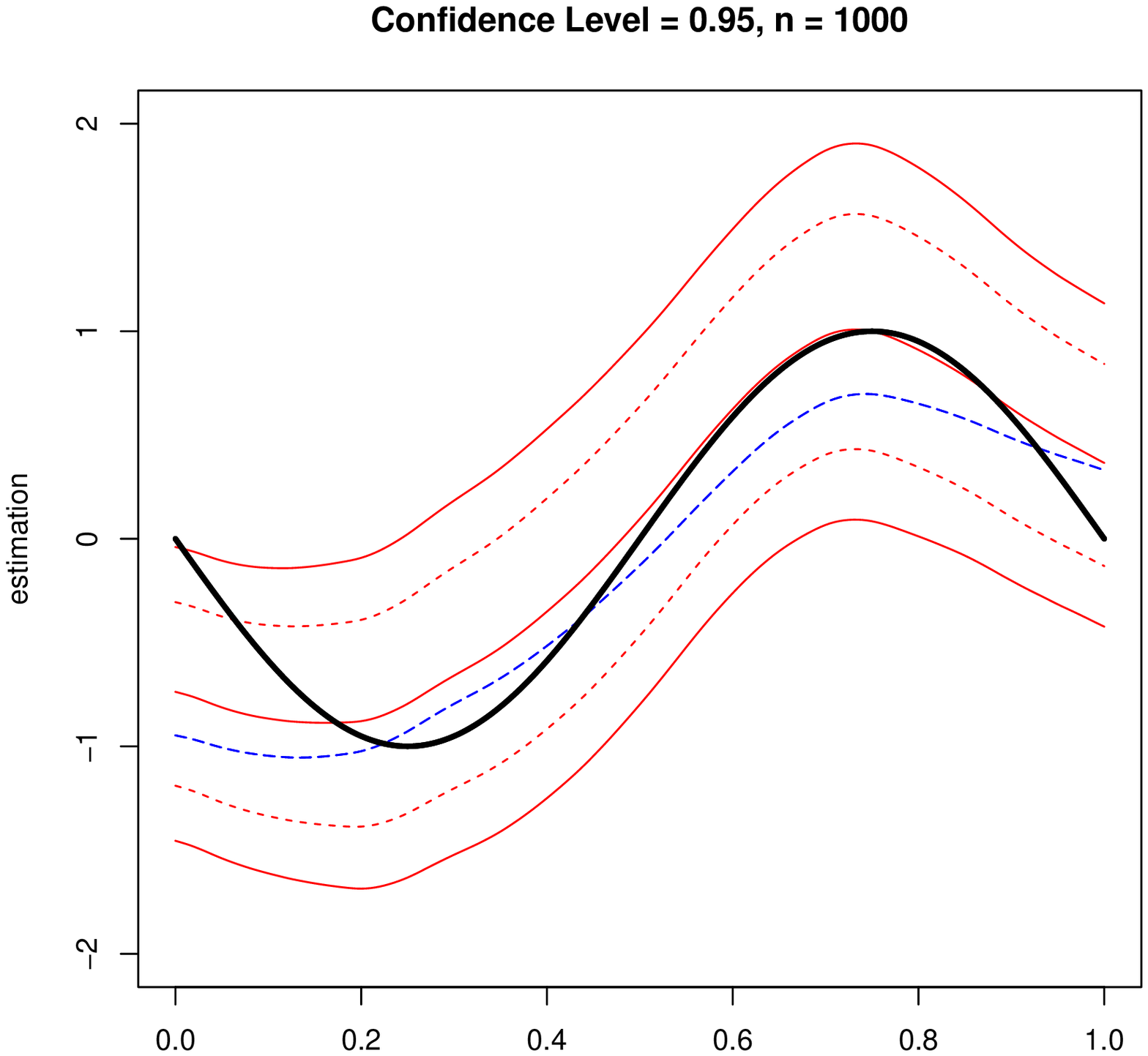}}
 \put(110,-370){(b)}
 \put(-250,-620){
 \includegraphics{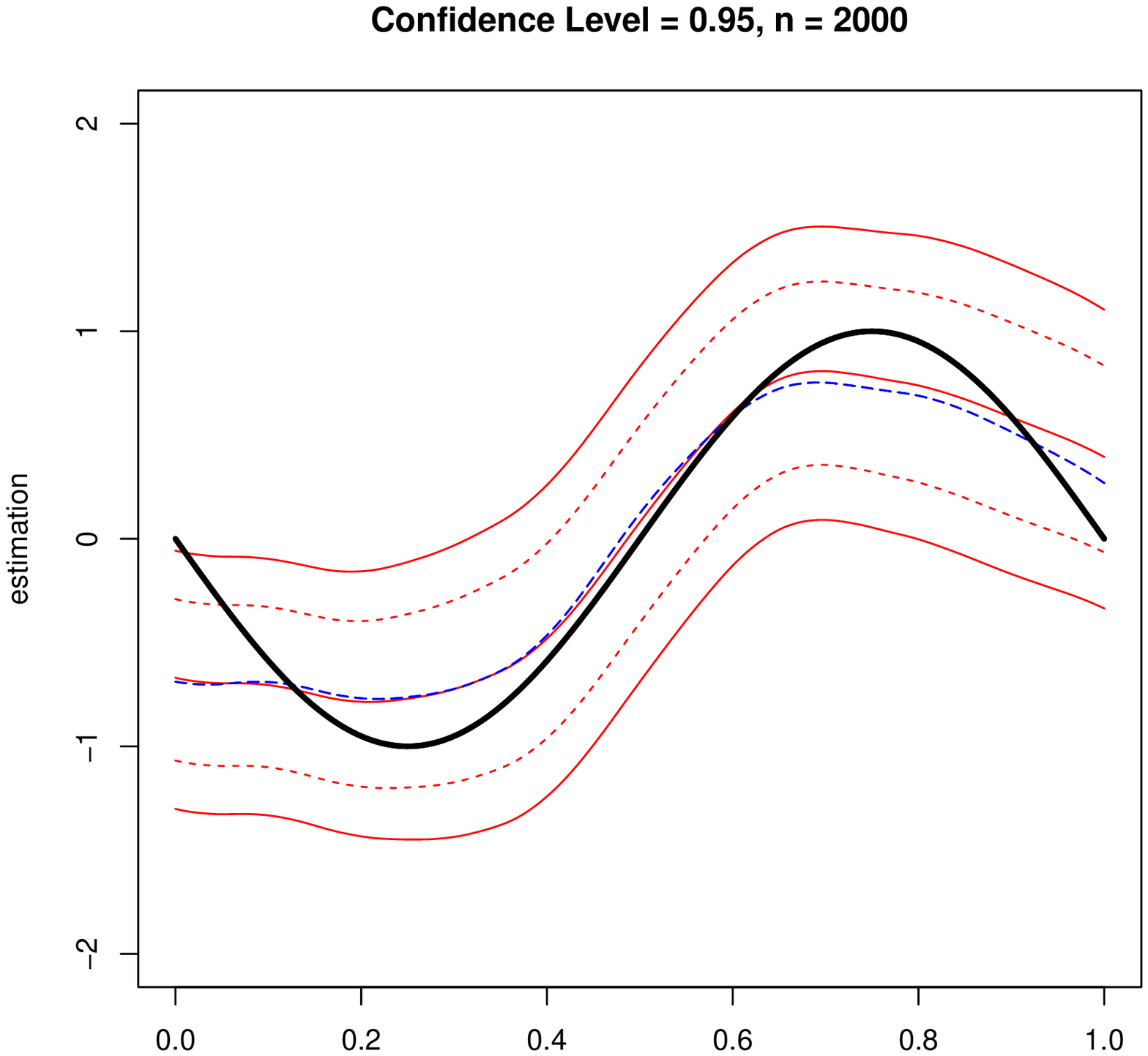}}
 \put(-130,-630){(c)}
 \put(-10,-620){
 \includegraphics{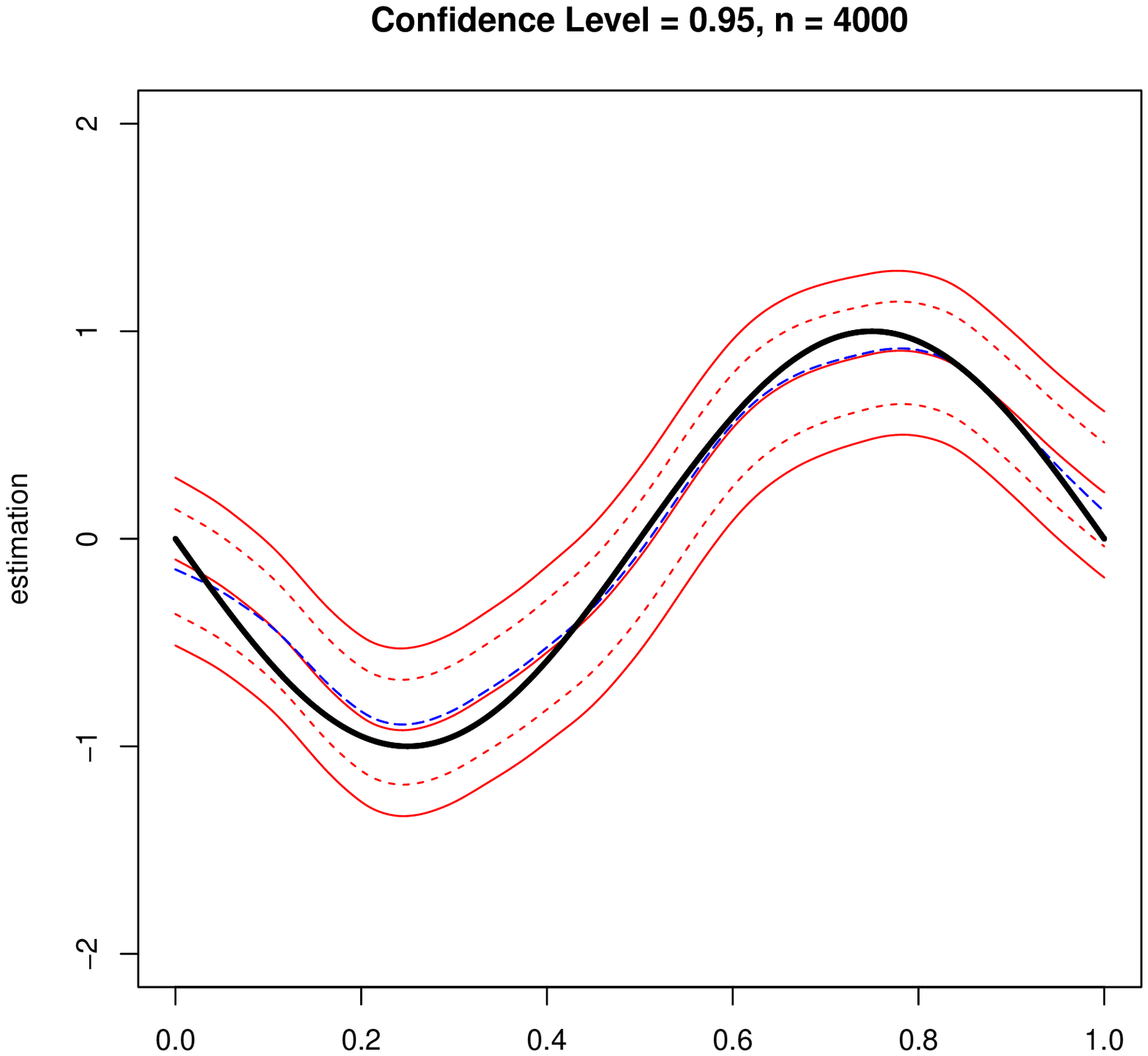}}
 \put(110,-630){(d)}
\end{picture}
\end{center}
\caption{Plots of $m_{1}(x_{1})$ - thick black line, $\tilde{m}_{\limfunc{K}%
,1}(x_{1})$ - blue dashed line, asymptotic $95\%$ pointwise confidence
intervals - red dashed line, $\hat{m}_{\func{SBK},1}(x_{1})$ and $95\%$
simultaneous confidence bands - red solid line for $r_{1}=0,r_{2}=0.5$ and
(a) $n=500$, (b) $n=1000$, (c) $n=2000$, (d) $n=4000$.}
\label{Figuresimu2}
\end{figure}

To have an impression of the actual function estimates, for $r_{1}=0$, $%
r_{2}=0.5$ with sample size $n=500$, $1000$, $2000$, $4000$, we have plotted
the SBK estimators and their 95\% asymptotic SCCs (red solid lines),
pointwise confidence intervals (red dashed lines), oracle estimators (blue
dashed lines) for the true functions $m_{1}$ (thick black lines) in Figure %
\ref{Figuresimu2}. Here we use $r_{1}=0$ because we want to give the 95\%
asymptotic SCCs, which need the observations be i.i.d to satisfy Assumption
(A5'). As expected by theoretical results, the estimation is closer to the
real function and the confidence band is narrower as sample size increasing. 

To compare the prediction performance of GAM\ and GAPLM, we introduce CAP
and AR first. For any score function $S$, one defines its alarm rate $%
F\left( s\right) =\Pr \left( S\leq s\right) $ and the hit rate $F_{\limfunc{D%
}}\left( s\right) =\Pr \left( S\leq s\left\vert \limfunc{D}\right. \right) $
where $\limfunc{D}$ represents the conditioning event of \textquotedblleft
default". Define the Cumulative Accuracy Profile ($\limfunc{CAP}$) curve as 
\begin{equation}
\limfunc{CAP}\left( u\right) =F_{\limfunc{D}}\left\{ F^{-1}\left( u\right)
\right\} ,u\in \left( 0,1\right) ,  \label{DEF:CAP}
\end{equation}%
which is the percentage of default-infected obligators that are found among
the first (according to their scores) $100u\%$ of all obligators. A perfect
rating method assigns all lowest scores to exactly the defaulters, so its
CAP curve linearly increases up and then stays at $1$, in other words, $%
\limfunc{CAP}_{\limfunc{P}}\left( u\right) =\mathrm{min} \left( u/p,1\right)
,u\in \left( 0,1\right) $, where $p$ denotes the unconditional default
probability. In contrast, a noninformative rating method with zero
discriminatory power displays a diagonal line $\limfunc{CAP}_{\limfunc{N}%
}\left( u\right) =u,u\in \left( 0,1\right) $. The CAP curve of a given
scoring method $S$ always locates between these two extremes and give
information about its performance.

The area between the CAP curve and the noninformative diagonal $\limfunc{CAP}%
_{\limfunc{N}}\left( u\right) \equiv u$ is $a_{R}$, whereas $a_{P}$ is the
area between the perfect CAP curve $\limfunc{CAP}_{\limfunc{P}}\left(
u\right) $ and the noninformative diagonal $\limfunc{CAP}_{\limfunc{N}%
}\left( u\right) $. Thus the CAP can be measured for example by Accuracy
Ratio (AR): the ratio of $a_{R}$ and $a_{P}$. 
\begin{equation*}
\limfunc{AR}=\frac{a_{R}}{a_{P}}=\frac{2\int_{0}^{1}\limfunc{CAP}\left(
u\right) du-1}{1-p},
\end{equation*}%
where $\limfunc{CAP}\left( u\right) $ is given in (\ref{DEF:CAP}). The AR
takes value in $\left[ 0,1\right] $, with value $0$ corresponding to the
noninformative scoring, and $1$ the perfect scoring method. A higher AR
indicates an overall higher discriminatory power of a method. Table \ref%
{Tablesimu3}\ shows the average and standard deviations of the ARs from $1000
$ replications using $k$-fold cross-validation with $k=2,10,100$ for $r_{1}=0
$, $r_{2}=0$ and $n=500$, $1000$, $2000$, $4000$. In each replication, we
randomly divide the set of observations into $k$ equal size folds and use th
rest $k-1$ folds as training data set to make prediction for each fold.
After we obtain all the prediction for each observation in the data set, we
compute the CAP\ and AR based on above formula. It is clear that GAPLM has
best predication accuracy. 
\begin{table}[th]
\begin{center}
\begin{tabular}{ccccc}
\hline\hline
$n$ &  & $k=2$ & $k=10$ & $k=100$ \\ \hline
$500$ & 
\begin{tabular}{c}
GLM \\ 
GAM \\ 
GAPLM%
\end{tabular}
& 
\begin{tabular}{c}
$0.6287\left( 0.0436\right) $ \\ 
$0.6222\left( 0.0732\right) $ \\ 
$0.6511\left( 0.0479\right) $%
\end{tabular}
& 
\begin{tabular}{c}
$0.6412\left( 0.0397\right) $ \\ 
$0.6706\left( 0.0393\right) $ \\ 
$0.6828\left( 0.0377\right) $%
\end{tabular}
& 
\begin{tabular}{c}
$0.6438\left( 0.0390\right) $ \\ 
$0.6756\left( 0.0400\right) $ \\ 
$0.6861\left( 0.0391\right) $%
\end{tabular}
\\ \hline
$1000$ & 
\begin{tabular}{c}
GLM \\ 
GAM \\ 
GAPLM%
\end{tabular}
& 
\begin{tabular}{c}
$0.6429\left( 0.0282\right) $ \\ 
$0.6735\left( 0.0438\right) $ \\ 
$0.6861\left( 0.0298\right) $%
\end{tabular}
& 
\begin{tabular}{c}
$0.6476\left( 0.0268\right) $ \\ 
$0.6863\left( 0.0326\right) $ \\ 
$0.6968\left( 0.0254\right) $%
\end{tabular}
& 
\begin{tabular}{c}
$0.6488\left( 0.0268\right) $ \\ 
$0.6929\left( 0.0261\right) $ \\ 
$0.7001\left( 0.0258\right) $%
\end{tabular}
\\ \hline
$2000$ & 
\begin{tabular}{c}
GLM \\ 
GAM \\ 
GAPLM%
\end{tabular}
& 
\begin{tabular}{c}
$0.6474\left( 0.0204\right) $ \\ 
$0.6842\left( 0.0615\right) $ \\ 
$0.6984\left( 0.0204\right) $%
\end{tabular}
& 
\begin{tabular}{c}
$0.6513\left( 0.0195\right) $ \\ 
$0.6984\left( 0.0286\right) $ \\ 
$0.7067\left( 0.0178\right) $%
\end{tabular}
& 
\begin{tabular}{c}
$0.6519\left( 0.0188\right) $ \\ 
$0.7000\left( 0.0185\right) $ \\ 
$0.7057\left( 0.0178\right) $%
\end{tabular}
\\ \hline
$4000$ & 
\begin{tabular}{c}
GLM \\ 
GAM \\ 
GAPLM%
\end{tabular}
& 
\begin{tabular}{c}
$0.6507\left( 0.0134\right) $ \\ 
$0.6889\left( 0.0243\right) $ \\ 
$0.7056\left( 0.0130\right) $%
\end{tabular}
& 
\begin{tabular}{c}
$0.6522\left( 0.0136\right) $ \\ 
$0.6968\left( 0.0403\right) $ \\ 
$0.7110\left( 0.0124\right) $%
\end{tabular}
& 
\begin{tabular}{c}
$0.6529\left( 0.0132\right) $ \\ 
$0.7079\left( 0.0164\right) $ \\ 
$0.7119\left( 0.0119\right) $%
\end{tabular}
\\ \hline
\end{tabular}%
\end{center}
\caption{The mean and standard deviation (in parentheses) of Accuracy Ratio
(AR) values for GLM, GAM, GAPLM for $r_{1}=0$, $r_{2}=0$ from 1000
replications. }
\label{Tablesimu3}
\end{table}

Last, to show the estimation performance when $\mathbf{T}$ has categorical
variables, we generate data using the same model above but add one more
categorical variable, i.e., $d_{1}=3$, $\mathbf{\beta }=\left( \beta
_{0},\beta _{1},\beta _{2},\beta _{3}\right) ^{\top}=\left( 1,1,1,1\right) ^{%
\T}$, $T_{3}=\left\{ 0,1\right\} $ with probability $0.5$ for $T_{3}=1$ and
independent with the other variables $T$ and $X$. Table \ref{Tablesimu4}
shows the bias, variances, MSEs and EFFs of $\hat{\beta}_{3}$ for $R=1000$
with sample sizes $\ n=500,1000,2000,4000$. The results show that the
estimator works as the asymptotic theory indicates. 
\begin{table}[th]
\begin{center}
\begin{tabular}{cccccc}
\hline\hline
$r$ & $n$ & $10\times \overline{\text{BIAS}}$ & $100\times \overline{\text{%
VARIANCE}}$ & $100\times \overline{\text{MSE}}$ & $\overline{\func{EFF}}%
\left( \hat{\beta}_{3}\right) $ \\ \hline
\multicolumn{1}{l}{$%
\begin{array}{c}
r_{1}=0 \\ 
r_{2}=0%
\end{array}%
$} & $%
\begin{array}{c}
500 \\ 
1000 \\ 
2000 \\ 
4000%
\end{array}%
$ & $%
\begin{array}{c}
1.476 \\ 
0.770 \\ 
0.448 \\ 
0.315%
\end{array}%
$ & $%
\begin{array}{c}
10.129 \\ 
4.437 \\ 
1.846 \\ 
0.937%
\end{array}%
$ & $%
\begin{array}{c}
12.309 \\ 
5.031 \\ 
2.047 \\ 
1.037%
\end{array}%
$ & $%
\begin{array}{c}
0.7634 \\ 
0.8343 \\ 
0.8929 \\ 
0.9572%
\end{array}%
$ \\ \hline
\multicolumn{1}{l}{$%
\begin{array}{c}
r_{1}=0.5 \\ 
\multicolumn{1}{l}{r_{2}=0}%
\end{array}%
$} & $%
\begin{array}{c}
500 \\ 
1000 \\ 
2000 \\ 
4000%
\end{array}%
$ & $%
\begin{array}{c}
1.336 \\ 
0.833 \\ 
0.423 \\ 
0.302%
\end{array}%
$ & $%
\begin{array}{c}
10.329 \\ 
4.221 \\ 
1.952 \\ 
0.944%
\end{array}%
$ & $%
\begin{array}{c}
12.115 \\ 
4.916 \\ 
2.132 \\ 
1.036%
\end{array}%
$ & $%
\begin{array}{c}
0.7445 \\ 
0.8267 \\ 
0.8832 \\ 
0.9436%
\end{array}%
$ \\ \hline
\multicolumn{1}{l}{$%
\begin{array}{l}
r_{1}=0 \\ 
\multicolumn{1}{c}{r_{2}=0.5}%
\end{array}%
$} & $%
\begin{array}{c}
500 \\ 
1000 \\ 
2000 \\ 
4000%
\end{array}%
$ & $%
\begin{array}{c}
1.441 \\ 
0.803 \\ 
0.489 \\ 
0.328%
\end{array}%
$ & $%
\begin{array}{c}
10.154 \\ 
4.446 \\ 
2.136 \\ 
0.924%
\end{array}%
$ & $%
\begin{array}{c}
12.231 \\ 
5.114 \\ 
2.376 \\ 
1.032%
\end{array}%
$ & $%
\begin{array}{c}
0.7556 \\ 
0.8430 \\ 
0.8785 \\ 
0.9572%
\end{array}%
$ \\ \hline
\multicolumn{1}{l}{$%
\begin{array}{c}
r_{1}=0.5 \\ 
r_{2}=0.5%
\end{array}%
$} & $%
\begin{array}{c}
500 \\ 
1000 \\ 
2000 \\ 
4000%
\end{array}%
$ & $%
\begin{array}{c}
1.475 \\ 
0.812 \\ 
0.524 \\ 
0.302%
\end{array}%
$ & $%
\begin{array}{c}
11.014 \\ 
4.464 \\ 
1.970 \\ 
0.966%
\end{array}%
$ & $%
\begin{array}{c}
13.190 \\ 
5.124 \\ 
2.245 \\ 
1.058%
\end{array}%
$ & $%
\begin{array}{c}
0.7794 \\ 
0.8314 \\ 
0.8852 \\ 
0.9529%
\end{array}%
$ \\ \hline
\end{tabular}%
\end{center}
\caption{The mean of $10\times $Bias, $100\times $Variances, $100\times $%
MSEs and EFFs of$\ \hat{\protect\beta}_{3}$ from 1000 replications. }
\label{Tablesimu4}
\end{table}

\subsection{\textbf{Example 2}}

The credit reform database, provided by the Research Data Center (RDC) of
the Humboldt Universit\"{a}t zu Berlin, was studied by using GAM model in
[11]. The data set contains $d=8$ financial ratios, which are shown in Table %
\ref{Tablevariable}, such as Operating\_Income/Total\_Assets and
log(Total\_Assets), of 18610 solvent ($Y=0$) and 1000 insolvent ($Y=1$)
German companies. The time period ranges from 1997 to 2002 and in the case
of the insolvent companies the information was gathered 2 years before the
insolvency took place. The last annual report of a company before it went
bankrupt receives the indicator $Y=1$ and for the rest (solvent) $Y=0$. In
the original data set, the variables are labeled as $Z_{\alpha }$. In order
to satisfy the Assumption (A4) in [11], we need the transformation: $%
X_{i\alpha }=F_{n\alpha }\left( Z_{i\alpha }\right) $, $\alpha =1,\ldots,8$,
where $F_{n\alpha }$ is the empirical cdf for the data $\left\{ X_{i\alpha
}\right\} _{i=1}^{n}$. See [4, 11] for more details of this data set. 
\begin{table}[th]
\begin{center}
\begin{tabular}{cccc}
\hline\hline
$\text{Ratio No.}$ & $\text{Definition}$ & $\text{Ratio No.}$ & $\text{%
Definition}$ \\ \hline
$Z_{1}$ & $\text{Net\_Income/Sales}$ & $Z_{5}$ & Cash/Total\_Assets \\ \hline
$Z_{2}$ & $\text{Operating\_Income/Total\_Assets}$ & $Z_{6}$ & $\text{%
Inventories/Sales}$ \\ \hline
$Z_{3}$ & $\text{Ebit/Total\_Assets}$ & $Z_{7}$ & $\text{Accounts%
\_Payable/Sales}$ \\ \hline
$Z_{4}$ & $\text{Total\_Liabilities/Total\_Assets}$ & $Z_{8}$ & $\text{%
log(Total\_Assets)}$ \\ \hline\hline
\end{tabular}%
\end{center}
\caption{Definitions of financial ratios.}
\label{Tablevariable}
\end{table}

\begin{figure}[tbh]
\begin{center}
\begin{picture}(-15,21)
 \put(-260,-370){
 \includegraphics{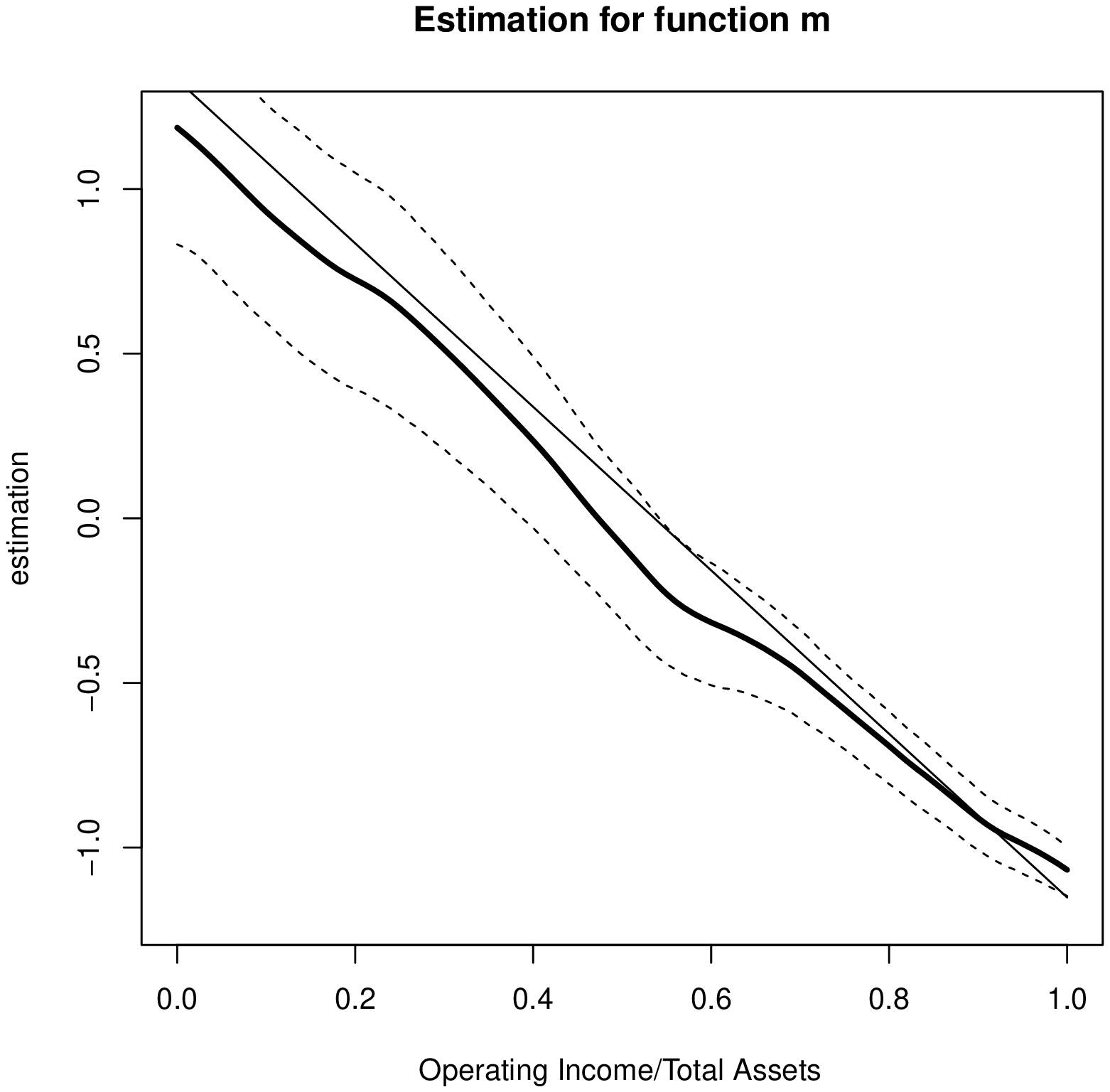}}
 \put(-140,-380){(a)}
 \put(-0,-370){
 \includegraphics{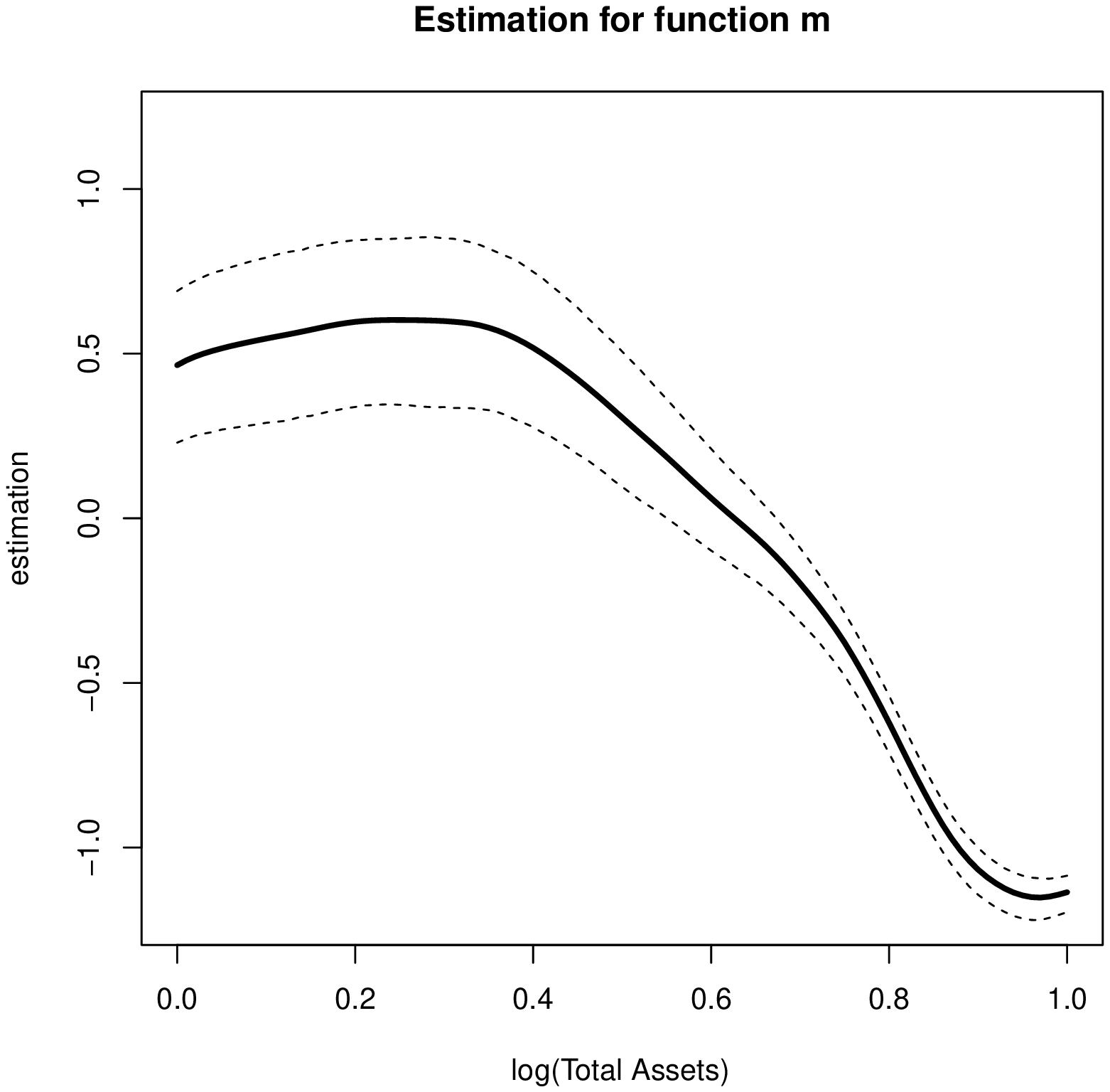}}
 \put(120,-380){(b)}
\end{picture}
\end{center}
\caption{Plots of estimations of component functions (a) $\hat{m}_{\func{SBK}%
,2}(x_{2})$ and (b) $\hat{m}_{\func{SBK},8}(x_{8})$ and asymptotic $95\%$
simultaneous confidence bands.}
\label{creditestimation}
\end{figure}

Using GAM and SBK method, we clearly see via the SCCs that the shape of $%
m_{2}\left( x_{2}\right) $ is linear. Figure \ref{creditestimation}(a) shows
that a linear line is covered by the SCCs of $\hat{m}_{2}$. We additionally
show the SCCs for another component function of log(Total\_Assets) in Figure %
\ref{creditestimation}(b). The SCCs do not cover a linear line. In fact,
among all the $8$ financial ratio considered, only $x_{2}$ yields a linear
influence. To improve the precision in statistical calibration and
interpretability, we can use GAPLM with parametric $m_{2}\left( x_{2}\right)
=\beta _{2}x_{2}$. 

For the RDC data, the in sample $\limfunc{AR}$ value obtained from GAPLM\ is 
$62.89\%$, which is very close to the $\limfunc{AR}$ value $63.05\%$
obtained from GAM in [11] and higher than the $\limfunc{AR}$ value $60.51\%$
obtained from SVM in [4]. To compare the prediction performance, we use the
AR introduced in Example 1. Then we randomly divide the data set into $k=2$,$%
10$ folds and obtain the prediction for each observation using the rest $k-1$
folds as training set. Based on the prediction of all the observation, we
can compute prediction AR value. Table \ref{Tableprediction} shows the mean
and standard deviation of the prediction AR values from $100$ replications.
GAPLM has higher prediction AR\ value than GAM for $99$ replications when $%
k=2$ and $100$ times when $k=10$. It is clear that GAPLM has best prediction
accuracy due to the better statistical calibration. 
\begin{table}[th]
\begin{center}
\begin{tabular}{ccc}
\hline\hline
& $k=2$ & $k=10$ \\ \hline
GLM & $0.5627\left( 0.0271\right) $ & $0.5751\left( 0.00162\right) $ \\ 
GAM & $0.5888\left( 0.0405\right) $ & $0.6123\left( 0.00219\right) $ \\ 
GAPLM & $0.5928\left( 0.0408\right) $ & $0.6164\left( 0.00196\right) $ \\ 
\hline
\end{tabular}%
\end{center}
\caption{The mean and standard deviation (in parentheses) of AR values for
GLM, GAM, GAPLM for $k$-fold Cross-validation with $k=2$ and $10$ from 1000
replications. }
\label{Tableprediction}
\end{table}

\section{Appendix}

\label{app}

\renewcommand{\theequation}{A.\arabic{equation}} \renewcommand{%
\thesubsection}{A.\arabic{subsection}} \renewcommand{\thelemma}{A.%
\arabic{lemma}} \setcounter{equation}{0} \setcounter{subsection}{0} %
\setcounter{lemma}{0}

\subsection{Preliminaries}

In the proofs that follow, we use \textquotedblleft $\CU$\textquotedblright\
and \textquotedblleft $\Cu$\textquotedblright\ to denote sequences of random
variables that are uniformly \textquotedblleft $\CO$\textquotedblright\ and
\textquotedblleft $\Co$ \textquotedblright\ of certain order. Denote the
theoretical inner product of $b_{J}$ and $1$ with respect to the $\alpha $%
-th marginal density $f_{\alpha }\left( x_{\alpha }\right) $ as $\
c_{J,\alpha }=\left\langle b_{J}\left( X_{\alpha }\right) ,1\right\rangle $ $%
=\int b_{J}\left( x_{\alpha }\right) f_{\alpha }\left( x_{\alpha }\right)
dx_{\alpha }$ and define the centered B spline basis $b_{J,\alpha }\left(
x_{\alpha }\right) $\ and the standardized B spline basis $B_{J,\alpha
}\left( x_{\alpha }\right) $ as 
\begin{equation*}
b_{J,\alpha }\left( x_{\alpha }\right) =b_{J}\left( x_{\alpha }\right) -%
\frac{c_{J,\alpha }}{c_{J-1,\alpha }}b_{J-1}\left( x_{\alpha }\right)
,B_{J,\alpha }\left( x_{\alpha }\right) =\frac{b_{J,\alpha }\left( x_{\alpha
}\right) }{\left\Vert b_{J,\alpha }\right\Vert _{2}},1\leq J\leq N+1,
\end{equation*}%
so that $\E B_{J,\alpha }\left( X_{\alpha }\right) =0$, $\E B_{J,\alpha
}^{2}\left( X_{\alpha }\right) =1$. Theorem A.2 in [20] shows that under
Assumptions (A1)-(A5) and (A7), constants $c_{0}\left( f\right) $, $C_{0}(f)$%
, $c_{1}\left( f\right) $ and $C_{1}(f)$ exist depending on the marginal
densities $f_{\alpha }\left( x_{\alpha }\right) ,1\leq \alpha \leq d,$ such
that $c_{0}\left( f\right) H\leq c_{J,\alpha }\leq C_{0}\left( f\right) H$, 
\begin{equation}
c_{1}\left( f\right) H\leq \left\Vert b_{J,\alpha }\right\Vert _{2}^{2}\leq
C_{1}(f)H.  \label{EQ: bJalpha}
\end{equation}

\begin{lemma}
\label{LEM:splineapprox}([1], p.149) For any $m\in C^{1}\left[ 0,1\right] $
with $m^{\prime }\in \limfunc{Lip}\left( \left[ 0,1\right] ,C_{\infty
}\right) $, there exist a constant $C_{\infty }>0$ and a function $g\in
G_{n}^{\left( 0\right) }\left[ 0,1\right] $ such that $\left\Vert
g-m\right\Vert _{\infty }\leq C_{\infty }H^{2}.$
\end{lemma}

\subsection{Oracle estimators\hfill}

\textsc{Proof of Theorem \ref{THM:betatilde-beta}. }(i) According to the
Mean Value Theorem, a vector $\mathbf{\bar{\beta}}$ between $\mathbf{\beta }$
and $\mathbf{\tilde{\beta}}$ exists such that $\left( \mathbf{\tilde{\beta}}-%
\mathbf{\beta }\right) \nabla ^{2}\tilde{\ell}_{\mathbf{\beta }}\left( 
\mathbf{\bar{\beta}}\right) =\nabla \tilde{\ell}_{\mathbf{\beta }}\left( 
\mathbf{\tilde{\beta}}\right) -\nabla \tilde{\ell}_{\mathbf{\beta }}\left( 
\mathbf{\beta }\right) =-\nabla \tilde{\ell}_{\mathbf{\beta }}\left( \mathbf{%
\beta }\right) $ since $\nabla \tilde{\ell}_{\mathbf{\beta }}\left( \mathbf{%
\tilde{\beta}}\right) =\mathbf{0}$, where 
\begin{equation*}
-\nabla ^{2}\tilde{\ell}_{\mathbf{\beta }}\left( \mathbf{\bar{\beta}}\right)
=n^{-1}\sum\nolimits_{i=1}^{n}b^{\prime \prime }\left\{ \mathbf{\bar{\beta}^{%
\T}T}_{i}+m\left( \mathbf{X}_{i}\right) \right\} \mathbf{T}_{i}\mathbf{T}%
_{i}^{\top }>c_{b}c_{\mathbf{Q}}\mathbf{I}_{d_{1}\times d_{1}}
\end{equation*}%
with $c_{b}>0$ according to (A2), and then the infeasible estimator is $%
\mathbf{\tilde{\beta}}=\func{argmax}_{\mathbf{a}\in \mathbb{R}^{1+d_{1}}}%
\tilde{\ell}_{\mathbf{\beta }}\left( \mathbf{a}\right) .$ 
\begin{equation*}
\nabla \tilde{\ell}_{\mathbf{\beta }}\left( \mathbf{\beta }\right)
=n^{-1}\tsum\nolimits_{i=1}^{n}\left[ Y_{i}\mathbf{T}_{i}-b^{\prime }\left\{ 
\mathbf{\beta ^{\top }T}_{i}+m\left( \mathbf{X}_{i}\right) \right\} \mathbf{T%
}_{i}\right] =n^{-1}\tsum\nolimits_{i=1}^{n}\sigma \left( \mathbf{T}_{i},%
\mathbf{X}_{i}\right) \varepsilon _{i}\mathbf{T}_{i}.
\end{equation*}%
We have $\left\vert n^{-1}\tsum\nolimits_{i=1}^{n}\sigma \left( \mathbf{T}%
_{i},\mathbf{X}_{i}\right) \varepsilon _{i}\mathbf{T}_{i}\right\vert =\CO%
_{a.s}\left( n^{-1/2}\log n\right) $ by Bernstein's Inequality as Lemma A.2
in [11], so 
\begin{equation*}
\left\vert \mathbf{\tilde{\beta}}-\mathbf{\beta }\right\vert =\CO%
_{a.s.}\left( n^{-1/2}\log n\right)
\end{equation*}%
according to $\mathbf{\tilde{\beta}}-\mathbf{\beta }=-\left\{ \nabla ^{2}%
\tilde{\ell}_{\mathbf{\beta }}\left( \mathbf{\bar{\beta}}\right) \right\}
^{-1}\nabla \tilde{\ell}_{\mathbf{\beta }}\left( \mathbf{\beta }\right) $.
Then%
\begin{equation*}
\nabla ^{2}\tilde{\ell}_{\mathbf{\beta }}\left( \mathbf{\bar{\beta}}\right) 
\overset{a.s.}{\rightarrow }\nabla ^{2}\tilde{\ell}_{\mathbf{\beta }}\left( 
\mathbf{\beta }\right) =-n^{-1}\sum_{i=1}^{n}b^{\prime \prime }\left\{ 
\mathbf{\beta ^{\top }T}_{i}+m\left( \mathbf{X}_{i}\right) \right\} \mathbf{T%
}_{i}\mathbf{T}_{i}^{\top },
\end{equation*}%
which converges to $-\E b^{\prime \prime }\left\{ m\left( \mathbf{T},\mathbf{%
X}\right) \right\} \mathbf{TT}^{\top }$ almost surely at the rate of $%
n^{-1/2}\log n$. So 
\begin{equation*}
\left\vert \mathbf{\tilde{\beta}}-\mathbf{\beta }-\left[ \E b^{\prime \prime
}\left\{ m\left( \mathbf{T},\mathbf{X}\right) \right\} \mathbf{TT}^{\top }%
\right] ^{-1}n^{-1}\tsum\nolimits_{i=1}^{n}\sigma \left( \mathbf{T}_{i},%
\mathbf{X}_{i}\right) \varepsilon _{i}\mathbf{T}_{i}\right\vert =\CO%
_{a.s.}\left( n^{-1}\left( \log n\right) ^{2}\right) .
\end{equation*}%
Since $n^{-1}\tsum\nolimits_{i=1}^{n}\sigma \left( \mathbf{T}_{i},\mathbf{X}%
_{i}\right) \varepsilon _{i}\mathbf{T}_{i}\overset{\tciLaplace }{\rightarrow 
}N\left( \mathbf{0},a\left( \phi \right) \left[ \E b^{\prime \prime }\left\{
m\left( \mathbf{T},\mathbf{X}\right) \right\} \mathbf{TT}^{\top }\right]
^{-1}\right) $ by central limit theorem, so Theorem \ref{THM:betatilde-beta}
(i) is proved by Slutsky's theorem.

(ii) The proof is trivial based on the properties of empirical likelihood
ratio for generalized linear model, see Theorem 3.2 in [15] and Corollary 1
in [7].\hfill $\square $

\subsection{Spline backfitted kernel estimators}

In this section, we present the proofs of Theorems \ref{THM:mhat-mtilde}, %
\ref{THM:bands} and \ref{THM:betahat-beta}. We write any $g\in G_{n}^{0}$ as 
$g=\mathbf{\lambda }^{\top}\mathbf{B}\left( \mathbf{X}_{i}\right) $ with
vector $\mathbf{\lambda }_{g}\mathbf{=}\left( \lambda _{J,\alpha }\right)
_{1\leq J\leq N+1,1\leq \alpha \leq d_{2}}^{\top}\in \mathbb{R}^{\left(
N+1\right) d_{2}}$ is the dimension of the additive spline space $G_{n}^{0}$%
, and%
\begin{equation*}
\mathbf{B}\left( \mathbf{x}\right) =\left\{ B_{1,1}\left( x_{1}\right)
,\ldots,B_{N+1,1}\left( x_{1}\right) ,\ldots,B_{1,d_{2}}\left(
x_{d_{2}}\right), \ldots,B_{N+1,d_{2}}\left( x_{d_{2}}\right) \right\}
^{\top}.
\end{equation*}%
Denote $\mathbf{B}\left( \mathbf{t},\mathbf{x}\right) =\left\{
1,t_{1},\ldots,t_{d_{1}},B_{1,1}\left( x_{1}\right) ,\ldots,B_{N+1,1}\left(
x_{1}\right) ,\ldots,B_{1,d_{2}}\left(
x_{d_{2}}\right),\ldots,B_{N+1,d_{2}}\left( x_{d_{2}}\right) \right\}
^{\top} $, \newline
$\mathbf{\lambda =}\left( \mathbf{\mathbf{\lambda }_{\beta }^{\top}},\mathbf{%
\mathbf{\lambda }_{g}^{\top}}\right) ^{\top}\mathbf{=}\left( \lambda
_{0},\lambda _{k},\lambda _{J,\alpha }\right) _{1\leq J\leq N+1,1\leq \alpha
\leq d_{2},1\leq k\leq d_{1}}^{\top}\in \mathbb{R}^{N_{d}}$ with $%
N_{d}=1+d_{1}+\left( N+1\right) d_{2}$ and 
\begin{equation*}
\hat{L}\left( \mathbf{\mathbf{\lambda }_{\beta },}g\right) =\hat{L}\left( 
\mathbf{\lambda }\right) =n^{-1}\sum_{i=1}^{n}\left[ Y_{i}\left\{ \mathbf{%
\lambda }^{\top}\mathbf{B}\left( \mathbf{T}_{i},\mathbf{X}_{i}\right)
\right\} -b\left\{ \mathbf{\lambda }^{\top}\mathbf{B}\left( \mathbf{T}_{i},%
\mathbf{X}_{i}\right) \right\} \right] ,
\end{equation*}%
which yields the gradient and Hessian formulae 
\begin{eqnarray*}
\nabla \hat{L}\left( \mathbf{\lambda }\right)
&=&n^{-1}\tsum\nolimits_{i=1}^{n}\left[ Y_{i}\mathbf{B}\left( \mathbf{T}_{i},%
\mathbf{X}_{i}\right) -b^{\prime }\left\{ \mathbf{\lambda }^{\top}\mathbf{B}%
\left( \mathbf{T}_{i},\mathbf{X}_{i}\right) \right\} \mathbf{B}\left( 
\mathbf{T}_{i},\mathbf{X}_{i}\right) \right] , \\
\nabla ^{2}\hat{L}\left( \mathbf{\lambda }\right)
&=&-n^{-1}\tsum\nolimits_{i=1}^{n}b^{\prime \prime }\left\{ \mathbf{\lambda }%
^{\top}\mathbf{B}\left( \mathbf{T}_{i},\mathbf{X}_{i}\right) \right\} 
\mathbf{B}\left( \mathbf{T}_{i},\mathbf{X}_{i}\right) \mathbf{B}\left( 
\mathbf{T}_{i},\mathbf{X}_{i}\right) ^{\top}.
\end{eqnarray*}%
The multivariate function $m\left( \mathbf{t,x}\right) $ is estimated by%
\begin{eqnarray*}
\hat{m}\left( \mathbf{t,x}\right) &=&\hat{\beta}_{0}+\tsum%
\nolimits_{k=1}^{d_{1}}\hat{\beta}_{k}t_{k}+\tsum\nolimits_{\alpha
=1}^{d_{2}}\hat{m}_{\alpha }\left( x_{\alpha }\right) =\mathbf{\hat{\lambda}}%
^{\top}\mathbf{B}\left( \mathbf{t},\mathbf{x}\right) , \\
\mathbf{\hat{\lambda}} &\mathbf{=}&\mathbf{\left( \mathbf{\mathbf{\hat{%
\lambda}}_{\beta }^{\top}},\mathbf{\mathbf{\hat{\lambda}}_{g}^{\top}}\right) 
}^{^{\top}}\mathbf{=}\left( \mathbf{\hat{\beta}}^{\top},\mathbf{\mathbf{\hat{%
\lambda}}_{g}^{\top}}\right) ^{\top}\mathbf{=}\left( \hat{\beta}_{k},\hat{%
\lambda}_{J,\alpha }\right) _{0\leq k\leq d_{1},1\leq \alpha \leq
d_{2},1\leq J\leq N+1}^{\top}=\func{argmax}_{\mathbf{\lambda }}\hat{L}\left( 
\mathbf{\lambda }\right) .
\end{eqnarray*}%
Lemma 14 of Stone (1986) ensures that with probability approaching $1$, $%
\mathbf{\hat{\lambda}}$ exists uniquely and that $\nabla \hat{L}\left( 
\mathbf{\hat{\lambda}}\right) \mathbf{=0}$. In addition, Lemma \ref%
{LEM:splineapprox} and (A1) provide a vector $\mathbf{\bar{\lambda}=}\left( 
\mathbf{\beta }^{\top},\mathbf{\mathbf{\bar{\lambda}}_{g}^{\top}}\right)
^{\top}$ and an additive spline function $\bar{m}$ such that 
\begin{equation}
\bar{m}\left( \mathbf{x}\right) =\mathbf{\mathbf{\bar{\lambda}}_{g}^{\top}B}%
\left( \mathbf{x}\right) ,\left\Vert \bar{m}-m\right\Vert _{\infty }\leq
C_{\infty }H^{2}.  \label{DEF: mbar lamdabar}
\end{equation}%
We first establish technical lemmas before proving Theorems \ref%
{THM:mhat-mtilde} and \ref{THM:betahat-beta}.

\begin{lemma}
\label{LEM:Ltildederiv}Under Assumptions (A1)-(A6) and (A8), as $%
n\rightarrow \infty $ 
\begin{eqnarray*}
\left\vert \nabla \hat{L}\left( \mathbf{\bar{\lambda}}\right) \right\vert &=&%
\CO_{a.s.}\left( H^{2}+n^{-1/2}\log n\right) , \\
\left\Vert \nabla \hat{L}\left( \mathbf{\bar{\lambda}}\right) \right\Vert &=&%
\CO_{a.s.}\left( H^{3/2}+H^{-1/2}n^{-1/2}\log n\right) .
\end{eqnarray*}
\end{lemma}

\noindent \textsc{Proof. }See supplement.$\hfill \square $\textsc{\ }

Define the following matrices: 
\begin{eqnarray*}
\mathbf{V} &=&\E\mathbf{B}\left( \mathbf{T,X}\right) \mathbf{B}\left( 
\mathbf{T,X}\right) ^{\top},\mathbf{S}=\mathbf{V}^{-1}, \\
\mathbf{V}_{n} &=&n^{-1}\tsum\nolimits_{i=1}^{n}\mathbf{B}\left( \mathbf{T}%
_{i},\mathbf{X}_{i}\right) \mathbf{B}\left( \mathbf{T}_{i},\mathbf{X}%
_{i}\right) ^{\top},\mathbf{S}_{n}=\mathbf{V}_{n}^{-1},
\end{eqnarray*}%
\begin{equation*}
\mathbf{V}_{b}=\E b^{\prime \prime }\left\{ m\left( \mathbf{T,X}\right)
\right\} \mathbf{B}\left( \mathbf{T,X}\right) \mathbf{B}\left( \mathbf{T,X}%
\right) ^{\top}=\left[ 
\begin{array}{ccc}
v_{b,00} & v_{b,0,k} & v_{b,0,J,\alpha } \\ 
v_{b,0,k^{\prime }} & v_{b,k,k^{\prime }} & v_{b,J,\alpha ,k^{\prime }} \\ 
v_{b,0,J^{\prime },\alpha ^{\prime }} & v_{b,J^{\prime },\alpha ^{\prime },k}
& v_{b,J,\alpha ,J^{\prime },\alpha ^{\prime }}%
\end{array}%
\right] _{N_{d}\times N_{d}}
\end{equation*}%
where $N_{d}=\left( N+1\right) d_{2}+1+d_{1}$, and%
\begin{equation}
\mathbf{S}_{b}=\mathbf{V}_{b}^{-1}=\left[ 
\begin{array}{ccc}
s_{b,00} & s_{b,0,k} & s_{b,0,J,\alpha } \\ 
s_{b,0,k^{\prime }} & s_{b,k,k^{\prime }} & s_{b,J,\alpha ,k^{\prime }} \\ 
s_{b,0,J^{\prime },\alpha ^{\prime }} & s_{b,J^{\prime },\alpha ^{\prime },k}
& _{b,J,\alpha ,J^{\prime },\alpha ^{\prime }}%
\end{array}%
\right] _{N_{d}\times N_{d}},  \label{DEF:VbSb}
\end{equation}%
For any vector $\mathbf{\lambda }\in \mathbb{R}^{N_{d}}$, denote%
\begin{equation*}
\mathbf{V}_{b}\left( \mathbf{\lambda }\right) =\E b^{\prime \prime }\left\{ 
\mathbf{\lambda }^{\top}\mathbf{B}\left( \mathbf{T,X}\right) \right\} 
\mathbf{B}\left( \mathbf{T,X}\right) \mathbf{B}\left( \mathbf{T,X}\right)
^{\top},\mathbf{S}_{b}\left( \mathbf{\lambda }\right) =\mathbf{V}%
_{b}^{-1}\left( \mathbf{\lambda }\right)
\end{equation*}%
\begin{equation}
\mathbf{V}_{n,b}\left( \mathbf{\lambda }\right) =-\nabla ^{2}\hat{L}\left( 
\mathbf{\lambda }\right) ,\mathbf{S}_{n,b}\left( \mathbf{\lambda }\right) =%
\mathbf{V}_{n,b}^{-1}\left( \mathbf{\lambda }\right) .  \label{DEF:VbSblamda}
\end{equation}

\begin{lemma}
\label{LEM:gamsinv}Under Assumptions (A2) and (A4),%
\begin{eqnarray*}
c_{\mathbf{V}}\mathbf{I}_{N_{d}} &\leq &\mathbf{V}\leq C_{\mathbf{V}}\mathbf{%
I}_{N_{d}},c_{\mathbf{S}}\mathbf{I}_{N_{d}}\leq \mathbf{S}\leq C_{\mathbf{S}}%
\mathbf{I}_{N_{d}}, \\
c_{\mathbf{V,}b}\mathbf{I}_{N_{d}} &\leq &\mathbf{V}_{b}\leq C_{\mathbf{V,}b}%
\mathbf{I}_{N_{d}},c_{\mathbf{S,}b}\mathbf{I}_{N_{d}}\leq \mathbf{S}_{b}\leq
C_{\mathbf{S,}b}\mathbf{I}_{N_{d}}.
\end{eqnarray*}%
Under Assumption (A2), (A4), (A5) and (A8), as $n\rightarrow \infty $ with
probability increasing to $1$%
\begin{eqnarray*}
c_{\mathbf{V}}\mathbf{I}_{N_{d}} &\leq &\mathbf{V}_{n}\left( \mathbf{\lambda 
}\right) \leq C_{\mathbf{V}}\mathbf{I}_{N_{d}},c_{\mathbf{S}}\mathbf{I}%
_{N_{d}}\leq \mathbf{S}_{n}\left( \mathbf{\lambda }\right) \leq C_{\mathbf{S}%
}\mathbf{I}_{N_{d}} \\
c_{\mathbf{V,}b}\mathbf{I}_{N_{d}} &\leq &\mathbf{V}_{n,b}\left( \mathbf{%
\lambda }\right) \leq C_{\mathbf{V,}b}\mathbf{I}_{N_{d}},c_{\mathbf{S,}b}%
\mathbf{I}_{N_{d}}\leq \mathbf{S}_{n,b}\left( \mathbf{\lambda }\right) \leq
C_{\mathbf{S,}b}\mathbf{I}_{N_{d}}\text{.}
\end{eqnarray*}
\end{lemma}

\noindent \textsc{Proof.} Using Lemma A.7 in [12] and boundness of function $%
b^{\prime }$. \hfill $\square $

Define three vectors $\mathbf{\Phi }_{b},\mathbf{\Phi }_{v},\mathbf{\Phi }%
_{r}$ as%
\begin{eqnarray*}
\mathbf{\Phi }_{b} &=&\left( \Phi _{b,J,\alpha }\right) _{0\leq k\leq
d_{1},1\leq \alpha \leq d_{2},1\leq J\leq N+1}^{\top} \\
&=&-\mathbf{S}_{b}n^{-1}\tsum\nolimits_{i=1}^{n}\left[ b^{\prime }\left\{
m\left( \mathbf{T}_{i},\mathbf{X}_{i}\right) \right\} -b^{\prime }\left\{ 
\bar{m}\left( \mathbf{T}_{i},\mathbf{X}_{i}\right) \right\} \right] \mathbf{B%
}\left( \mathbf{T}_{i},\mathbf{X}_{i}\right) ,
\end{eqnarray*}
\begin{eqnarray*}
\mathbf{\Phi }_{v} &=&\left( \Phi _{v,J,\alpha }\right) _{0\leq k\leq
d_{1},1\leq \alpha \leq d_{2},1\leq J\leq N+1}^{\top} \\
&=&-\mathbf{S}_{b}n^{-1}\tsum\nolimits_{i=1}^{n}\left[ \sigma \left( \mathbf{%
T}_{i},\mathbf{X}_{i}\right) \varepsilon _{i}\right] \mathbf{B}\left( 
\mathbf{T}_{i},\mathbf{X}_{i}\right) , \\
\mathbf{\Phi }_{r} &=&\left( \Phi _{r,J,\alpha }\right) _{0\leq k\leq
d_{1},1\leq \alpha \leq d_{2},1\leq J\leq N+1}^{\top} \\
&=&\mathbf{\mathbf{\hat{\lambda}}-\mathbf{\bar{\lambda}}-\Phi }_{b}-\mathbf{%
\Phi }_{v}.
\end{eqnarray*}

\begin{lemma}
\label{LEM: lamda phai}Under Assumptions (A1)-(A6) and (A8), as $%
n\rightarrow \infty $%
\begin{eqnarray}
\left\Vert \mathbf{\hat{\lambda}}-\mathbf{\bar{\lambda}}\right\Vert &=&\CO%
_{a.s.}\left( H^{3/2}+H^{-1/2}n^{-1/2}\log n\right) ,
\label{EQ:lamdahat-lamdabar} \\
\left\Vert \mathbf{\Phi }_{r}\right\Vert &=&\CO_{p}\left( H^{-3/2}n^{-1}\log
n\right) ,  \label{EQ:Phi_rbd} \\
\left\Vert \mathbf{\Phi }_{b}\right\Vert &=&\CO_{a.s.}\left( H^{2}\right)
,\left\Vert \mathbf{\Phi }_{v}\right\Vert =\CO_{a.s.}\left(
H^{-1/2}n^{-1/2}\log n\right) .  \notag
\end{eqnarray}
\end{lemma}

\noindent \textsc{Proof. }See supplement.$\hfill \square $

\begin{lemma}
\label{LEM:mhat-m}Under Assumptions (A1)-(A6) and (A8), as $n\rightarrow
\infty $%
\begin{eqnarray*}
\left\Vert \hat{m}-\bar{m}\right\Vert _{2,n}+\left\Vert \hat{m}-\bar{m}%
\right\Vert _{2} &=&\CO_{a.s.}\left( H^{3/2}+H^{-1/2}n^{-1/2}\log n\right) ,
\\
\left\Vert \hat{m}-m\right\Vert _{2,n}+\left\Vert \hat{m}-m\right\Vert _{2}
&=&\CO_{a.s.}\left( H^{3/2}+H^{-1/2}n^{-1/2}\log n\right) .
\end{eqnarray*}
\end{lemma}

\noindent \textsc{Proof.} Lemma \ref{LEM:gamsinv} implies 
\begin{eqnarray*}
\left\Vert \hat{m}-\bar{m}\right\Vert _{2,n}+\left\Vert \hat{m}-\bar{m}%
\right\Vert _{2} &\leq &2C_{\mathbf{V}}\left\Vert \mathbf{\mathbf{\hat{%
\lambda}}}_{\mathbf{g}}-\mathbf{\mathbf{\bar{\lambda}}_{g}}\right\Vert \\
&=&\CO_{a.s.}\left( H^{3/2}+H^{-1/2}n^{-1/2}\log n\right) .
\end{eqnarray*}%
The Lemma follows $\left\Vert \bar{m}-m\right\Vert _{\infty }+\left\Vert 
\bar{m}-m\right\Vert _{2}+\left\Vert \bar{m}-m\right\Vert _{2,n}=\CO\left(
H^{2}\right) $ by (\ref{DEF: mbar lamdabar}). \hfill \hfill \hfill $\square $

\noindent \textsc{Proof of Theorem \ref{THM:mhat-mtilde}.} According to (\ref%
{DEF:lhat}) and the Mean Value Theorem, a $\bar{m}_{\limfunc{K},1}\left(
x_{1}\right) $ between $\hat{m}_{\func{SBK},1}\left( x_{1}\right) $ and $%
\tilde{m}_{\limfunc{K},1}\left( x_{1}\right) $ exists such that%
\begin{equation*}
\hat{\ell}_{m_{1}}^{\prime }\left\{ \hat{m}_{\func{SBK},1}\left(
x_{1}\right) ,x_{1}\right\} -\hat{\ell}^{\prime }\left\{ \tilde{m}_{\limfunc{%
K},1}\left( x_{1}\right) ,x_{1}\right\} =\hat{\ell}_{m_{1}}^{\prime \prime
}\left( \bar{m}_{\limfunc{K},1}\left( x_{1}\right) ,x_{1}\right) \left\{ 
\hat{m}_{\func{SBK},1}\left( x_{1}\right) -\tilde{m}_{\limfunc{K},1}\left(
x_{1}\right) \right\} ,
\end{equation*}%
Then according to $\hat{\ell}_{m_{1}}^{\prime }\left\{ \hat{m}_{\func{SBK}%
,1}\left( x_{1}\right) ,x_{1}\right\} =0$, one has 
\begin{equation*}
\hat{m}_{\func{SBK},1}\left( x_{1}\right) -\tilde{m}_{\limfunc{K},1}\left(
x_{1}\right) =-\frac{\hat{\ell}_{m_{1}}^{\prime }\left\{ \tilde{m}_{\limfunc{%
K},1}\left( x_{1}\right) ,x_{1}\right\} }{\hat{\ell}_{m_{1}}^{\prime \prime
}\left\{ \bar{m}_{\limfunc{K},1}\left( x_{1}\right) ,x_{1}\right\} }.
\end{equation*}%
The theorem then follows Lemmas A.15 and A.16 in [11] with small
modification including variable $\mathbf{T}$.

\noindent \textsc{Proof of Theorem \ref{THM:bands}.} It follows Theorem \ref%
{THM:mhat-mtilde} and the same proof of Theorem 1 in [25].\hfill $\square $

\noindent \textsc{Proof of Theorem \ref{THM:betahat-beta}. }See supplement.$%
\hfill \square $

\section{ACKNOWLEDGEMENTS}

Financial support from the Deutsche Forschungsgemeinschaft (DFG) via SFB 649
\textquotedblleft Economic Risk\textquotedblright , and International
Research Training Group (IRTG) 1792 are gratefully acknowledged. The authors
thank the Associate Editor and two referees for their comments and
suggestions which have led to substantial improvement of this work.

\section{SUPPLEMENTARY MATERIALS}

\noindent \textbf{Supplement to \textquotedblleft Statistical Inference for
Generalized Additive Partially Linear Model\textquotedblright }: Supplement
containing theoretical proof of Lemmas A.2, A.4 and Theorem 4 referenced in
the main article.

\noindent \textbf{gaplmsbk.R}: R-package containing code to perform SBK
estimation for component functions in generalized additive partially linear
model available on https://github.com.

\newpage \setcounter{page}{1}

\begin{center}
{\large Supplement to \textquotedblleft Statistical Inference for
Generalized Additive Partially Linear Model\textquotedblright}

\bigskip
\end{center}

\noindent \textbf{Proof of Lemma A.2 }\textsc{\ }%
\begin{eqnarray*}
\nabla \hat{L}\left( \mathbf{\bar{\lambda}}\right)
&=&n^{-1}\tsum\nolimits_{i=1}^{n}\left[ Y_{i}\mathbf{B}\left( \mathbf{T}_{i},%
\mathbf{X}_{i}\right) -b^{\prime }\left\{ \mathbf{\bar{\lambda}}^{\top }%
\mathbf{B}\left( \mathbf{T}_{i},\mathbf{X}_{i}\right) \right\} \mathbf{B}%
\left( \mathbf{T}_{i},\mathbf{X}_{i}\right) \right] \\
&=&n^{-1}\tsum\nolimits_{i=1}^{n}\left[ b^{\prime }\left\{ m\left( \mathbf{T}%
_{i},\mathbf{X}_{i}\right) \right\} -b^{\prime }\left\{ \bar{m}\left( 
\mathbf{T}_{i},\mathbf{X}_{i}\right) \right\} +\sigma \left( \mathbf{X}%
_{i}\right) \varepsilon _{i}\right] \mathbf{B}\left( \mathbf{T}_{i},\mathbf{X%
}_{i}\right)
\end{eqnarray*}%
The first $\left( 1+d_{1}\right) $ elements of the above vector is 
\begin{equation*}
n^{-1}\sum_{i=1}^{n}\left[ \left[ b^{\prime }\left\{ m\left( \mathbf{T}_{i},%
\mathbf{X}_{i}\right) \right\} -b^{\prime }\left\{ \bar{m}\left( \mathbf{T}%
_{i},\mathbf{X}_{i}\right) \right\} \right] +\sigma \left( \mathbf{X}%
_{i}\right) \varepsilon _{i}\right] T_{ik},0\leq k\leq d_{1},
\end{equation*}%
with $T_{i0}=1$. These elements are $\CO_{a.s.}\left( H^{2}+n^{-1/2}\log
n\right) $ according to (\ref{DEF: mbar lamdabar}). The other elements can
be written as%
\begin{equation*}
n^{-1}\tsum\nolimits_{i=1}^{n}\left[ \xi _{i,J,\alpha ,n}+\E\left[ b^{\prime
}\left\{ m\left( \mathbf{T}_{i},\mathbf{X}_{i}\right) \right\} -b^{\prime
}\left\{ \bar{m}\left( \mathbf{T}_{i},\mathbf{X}_{i}\right) \right\} \right]
B_{J,\alpha }\left( X_{i\alpha }\right) +\sigma \left( \mathbf{X}_{i}\right)
\varepsilon _{i}B_{J,\alpha }\left( X_{i\alpha }\right) \right] ,
\end{equation*}%
where $\xi _{i,J,\alpha ,n}$ is%
\begin{equation*}
\left[ b^{\prime }\left\{ m\left( \mathbf{T}_{i},\mathbf{X}_{i}\right)
\right\} -b^{\prime }\left\{ \bar{m}\left( \mathbf{T}_{i},\mathbf{X}%
_{i}\right) \right\} \right] B_{J,\alpha }\left( X_{i\alpha }\right) -\E%
\left[ \left[ b^{\prime }\left\{ m\left( \mathbf{T}_{i},\mathbf{X}%
_{i}\right) \right\} -b^{\prime }\left\{ \bar{m}\left( \mathbf{T}_{i},%
\mathbf{X}_{i}\right) \right\} \right] B_{J,\alpha }\left( X_{i\alpha
}\right) \right] .
\end{equation*}%
According to (\ref{EQ: bJalpha}) and (\ref{DEF: mbar lamdabar}), one has%
\begin{align*}
& \left\vert \E\left[ b^{\prime }\left\{ m\left( \mathbf{T}_{i},\mathbf{X}%
_{i}\right) \right\} -b^{\prime }\left\{ \bar{m}\left( \mathbf{T}_{i},%
\mathbf{X}_{i}\right) \right\} \right] B_{J,\alpha }\left( X_{i\alpha
}\right) \right\vert \\
& \leq \E\left\vert b^{\prime }\left\{ m\left( \mathbf{T}_{i},\mathbf{X}%
_{i}\right) \right\} -b^{\prime }\left\{ \bar{m}\left( \mathbf{T}_{i},%
\mathbf{X}_{i}\right) \right\} \right\vert \frac{\left\vert b_{J,\alpha
}\left( X_{i\alpha }\right) \right\vert }{\left\Vert b_{J,\alpha
}\right\Vert _{2}} \\
& \leq c\left\Vert m-\bar{m}\right\Vert _{\infty }\mathrm{max}_{1\leq J\leq
N+1,1\leq \alpha \leq d_{2}}\left\Vert b_{J,\alpha }\right\Vert _{2}^{-1}%
\mathrm{max}_{1\leq J\leq N+1,1\leq \alpha \leq d_{2}}\E\left\vert
b_{J,\alpha }\left( X_{i\alpha }\right) \right\vert \\
& =\CO\left( H^{2}\times H^{-1/2}\times H\right) =\CO\left( H^{5/2}\right) ,
\end{align*}%
for some constant $c$ and likewise for any $p\geq 2$%
\begin{align*}
& \E\left\vert b^{\prime }\left\{ m\left( \mathbf{T}_{i},\mathbf{X}%
_{i}\right) \right\} -b^{\prime }\left\{ \bar{m}\left( \mathbf{T}_{i},%
\mathbf{X}_{i}\right) \right\} \right\vert ^{p}\left\vert B_{J,\alpha
}\left( X_{i\alpha }\right) \right\vert ^{p} \\
& \leq \left( cH^{5/2}\right) ^{p-2}\E\left\vert b^{\prime }\left\{ m\left( 
\mathbf{T}_{i},\mathbf{X}_{i}\right) \right\} -b^{\prime }\left\{ \bar{m}%
\left( \mathbf{T}_{i},\mathbf{X}_{i}\right) \right\} \right\vert ^{2}\frac{%
b_{J,\alpha }^{2}\left( X_{i\alpha }\right) }{\left\Vert b_{J,\alpha
}\right\Vert _{2}^{2}},
\end{align*}%
and%
\begin{align*}
& \E\left[ b^{\prime }\left\{ m\left( \mathbf{T}_{i},\mathbf{X}_{i}\right)
\right\} -b^{\prime }\left\{ \bar{m}\left( X_{i\alpha }\right) \right\} %
\right] ^{2}B_{J,\alpha }^{2}\left( X_{i\alpha }\right) \\
& \leq c\left\Vert m-\bar{m}\right\Vert _{\infty }^{2}\mathrm{max}_{1\leq
J\leq N+1,1\leq \alpha \leq d_{2}}\left\Vert b_{J,\alpha }\right\Vert
_{2}^{-2}\mathrm{max}_{1\leq J\leq N+1,1\leq \alpha \leq d_{2}}\E\left\vert
b_{J,\alpha }^{2}\left( X_{i\alpha }\right) \right\vert =\CO\left(
H^{4}\right) .
\end{align*}%
Using these bounds and applying Lemma A.2 in [11], one has $\left\vert
n^{-1}\sum_{i=1}^{n}\xi _{i,J,\alpha ,n}\right\vert =\CO_{a.s.}\left(
H^{2}n^{-1/2}\log n\right) $ and 
\begin{equation*}
n^{-1}\left\vert \tsum\nolimits_{i=1}^{n}\sigma \left( \mathbf{X}_{i}\right)
\varepsilon _{i}B_{J,\alpha }\left( X_{i\alpha }\right) \right\vert =\CO%
_{a.s.}\left( n^{-1/2}\log n\right) .
\end{equation*}%
The lemma is then proved.\hfill \hfill \hfill $\square $

\bigskip

\noindent \textbf{Proof of Lemma A.4 }The Mean Value Theorem implies that an 
$N_{d}\times N_{d}$ diagonal matrix $\mathbf{t}$ exists whose diagonal
elements are in $\left[ 0,1\right] $, such that for $\mathbf{\hat{\lambda}}%
^{\ast }=\mathbf{t\hat{\lambda}+}\left( \mathbf{I}_{N_{d}}-\mathbf{t}\right) 
\mathbf{\bar{\lambda}}$%
\begin{equation*}
\nabla \hat{L}\left( \mathbf{\hat{\lambda}}\right) -\nabla \hat{L}\left( 
\mathbf{\bar{\lambda}}\right) =\nabla ^{2}\hat{L}\left( \mathbf{\hat{\lambda}%
}^{\ast }\right) \left( \mathbf{\hat{\lambda}-\bar{\lambda}}\right) .
\end{equation*}%
Since, as noted before, that $\nabla \hat{L}\left( \mathbf{\hat{\lambda}}%
\right) =\mathbf{0}$, the above equation becomes%
\begin{equation*}
\mathbf{\hat{\lambda}-\bar{\lambda}}=-\left\{ \nabla ^{2}\hat{L}\left( 
\mathbf{\hat{\lambda}}^{\ast }\right) \right\} ^{-1}\nabla \hat{L}\left( 
\mathbf{\bar{\lambda}}\right) .
\end{equation*}%
According to (\ref{DEF:VbSblamda}), 
\begin{equation*}
-\nabla ^{2}\hat{L}\left( \mathbf{\lambda }\right)
=n^{-1}\tsum\nolimits_{i=1}^{n}b^{\prime \prime }\left\{ \mathbf{\lambda }^{%
\T}\mathbf{B}\left( \mathbf{T}_{i},\mathbf{X}_{i}\right) \right\} \mathbf{B}%
\left( \mathbf{T}_{i},\mathbf{X}_{i}\right) \mathbf{B}\left( \mathbf{T}_{i},%
\mathbf{X}_{i}\right) ^{\top}=\mathbf{V}_{n,b}\left( \mathbf{\lambda }%
\right) ,
\end{equation*}%
Lemma \ref{LEM:gamsinv} implies that with probability approaching $1$%
\begin{equation*}
c_{\mathbf{V},b}\mathbf{I}_{N_{d}}\leq -\nabla ^{2}\hat{L}\left( \mathbf{%
\hat{\lambda}}^{\ast }\right) \leq C_{\mathbf{V},b}\mathbf{I}_{N_{d}}\text{.}
\end{equation*}%
Then (\ref{EQ:lamdahat-lamdabar}) follows Lemma \ref{LEM:Ltildederiv}.
Furthermore, $\left\Vert \mathbf{\hat{\lambda}}^{\ast }-\mathbf{\bar{\lambda}%
}\right\Vert =\CO_{a.s}\left( H^{3/2}+H^{-1/2}n^{-1/2}\log n\right) $ as
well according to $\mathbf{\hat{\lambda}}^{\ast }$'s definition. Note that
Taylor expansion ensures that for any vector $\mathbf{a}\in \mathbb{R}%
^{N_{d}}$%
\begin{equation*}
\mathbf{a}^{\top}\left\{ \nabla ^{2}\hat{L}\left( \mathbf{\hat{\lambda}}%
^{\ast }\right) -\nabla ^{2}\hat{L}\left( \mathbf{\bar{\lambda}}\right)
\right\} \mathbf{a\leq }\left\Vert b^{\prime \prime \prime }\right\Vert
_{\infty }\mathrm{max}_{1\leq i\leq n}\left\vert \mathbf{\hat{\lambda}}%
^{\ast \T}\mathbf{B}\left( \mathbf{T}_{i},\mathbf{X}_{i}\right) -\mathbf{%
\bar{\lambda}}^{\top}\mathbf{B}\left( \mathbf{T}_{i},\mathbf{X}_{i}\right)
\right\vert \mathbf{a}^{\top}\mathbf{V}_{n}\mathbf{a}
\end{equation*}%
while by Cauchy Schwartz inequality%
\begin{align*}
& \mathrm{max}_{1\leq i\leq n}\left\vert \mathbf{\hat{\lambda}}^{\ast \T}%
\mathbf{B}\left( \mathbf{T}_{i},\mathbf{X}_{i}\right) -\mathbf{\bar{\lambda}}%
^{\top}\mathbf{B}\left( \mathbf{T}_{i},\mathbf{X}_{i}\right) \right\vert
\leq \left\Vert \mathbf{\hat{\lambda}}^{\ast }-\mathbf{\bar{\lambda}}%
\right\Vert \mathrm{max}_{1\leq i\leq n}\left\Vert \mathbf{B}\left( \mathbf{T%
}_{i},\mathbf{X}_{i}\right) \right\Vert \\
& =\CO_{a.s.}\left( H^{3/2}+H^{-1/2}n^{-1/2}\log n\right) \times \CO%
_{p}\left( H^{-1/2}\right) =\CO_{p}\left( H+H^{-1}n^{-1/2}\log n\right) .
\end{align*}%
Consequently, one has the following bound on the difference of two Hessian
matrices%
\begin{equation*}
\mathrm{sup}_{\mathbf{a}\in R^{N_{d}}}\left\Vert \left( \nabla ^{2}\hat{L}%
\left( \mathbf{\hat{\lambda}}^{\ast }\right) -\nabla ^{2}\hat{L}\left( 
\mathbf{\bar{\lambda}}\right) \right) \mathbf{a}\right\Vert \left\Vert 
\mathbf{a}\right\Vert ^{-1}=\CO_{p}\left( H+H^{-1}n^{-1/2}\log n\right) .
\end{equation*}%
Denote next%
\begin{eqnarray*}
\mathbf{\hat{d}} &=&-\left\{ \nabla ^{2}\hat{L}\left( \mathbf{\hat{\lambda}}%
^{\ast }\right) \right\} ^{-1}\nabla \hat{L}\left( \mathbf{\bar{\lambda}}%
\right) =\mathbf{\hat{\lambda}-\bar{\lambda}} \\
\mathbf{\bar{d}} &=&-\left\{ \nabla ^{2}\hat{L}\left( \mathbf{\bar{\lambda}}%
\right) \right\} ^{-1}\nabla \hat{L}\left( \mathbf{\bar{\lambda}}\right)
\end{eqnarray*}%
then $\left\Vert \mathbf{\hat{d}}\right\Vert =\CO_{a.s.}\left(
H^{3/2}+H^{-1/2}n^{-1/2}\log n\right) $ and so is $\left\Vert \mathbf{\bar{d}%
}\right\Vert $ by similar arguments. Furthermore,%
\begin{equation*}
\nabla ^{2}\hat{L}\left( \mathbf{\hat{\lambda}}^{\ast }\right) \left( 
\mathbf{\hat{d}-\bar{d}}\right) \mathbf{=}\left\{ \nabla ^{2}\hat{L}\left( 
\mathbf{\bar{\lambda}}\right) -\nabla ^{2}\hat{L}\left( \mathbf{\hat{\lambda}%
}^{\ast }\right) \right\} \mathbf{\bar{d}}
\end{equation*}%
entails that%
\begin{eqnarray*}
\left\Vert \mathbf{\hat{d}-\bar{d}}\right\Vert &=&\CO_{a.s.}\left(
H^{3/2}+H^{-1/2}n^{-1/2}\log n\right) \times \CO_{p}\left(
H+H^{-1}n^{-1/2}\log n\right) \\
&=&\CO_{p}\left( H^{5/2}+H^{-3/2}n^{-1}\log ^{2}n\right) .
\end{eqnarray*}%
Denote%
\begin{equation*}
\mathbf{\tilde{d}}=\left[ n^{-1}\tsum\nolimits_{i=1}^{n}b^{\prime \prime
}\left\{ m\left( \mathbf{T}_{i},\mathbf{X}_{i}\right) \right\} \mathbf{B}%
\left( \mathbf{T}_{i},\mathbf{X}_{i}\right) \mathbf{B}\left( \mathbf{T}_{i},%
\mathbf{X}_{i}\right) ^{\top}\right] ^{-1}\nabla \hat{L}\left( \mathbf{\bar{%
\lambda}}\right) .
\end{equation*}%
Using similar calculations, one can show that%
\begin{eqnarray*}
\left\Vert \mathbf{\tilde{d}-\bar{d}}\right\Vert &=&\CO_{a.s.}\left(
H^{3/2}+H^{-1/2}n^{-1/2}\log n\right) \times \CO_{a.s}\left( H^{2}\right) \\
&=&\CO_{a.s.}\left( H^{7/2}+H^{3/2}n^{-1/2}\log n\right) ,
\end{eqnarray*}%
\begin{eqnarray*}
\left\Vert \mathbf{\tilde{d}-\Phi }_{b}-\mathbf{\Phi }_{v}\right\Vert &=&\CO%
_{a.s.}\left( H^{3/2}+H^{-1/2}n^{-1/2}\log n\right) \times \CO_{a.s}\left(
H^{-1/2}n^{-1/2}\log n\right) \\
&=&\CO_{a.s.}\left( Hn^{-1/2}\log n+H^{-1}n^{-1}\log ^{2}n\right) ,
\end{eqnarray*}%
Putting together the above proves (\ref{EQ:Phi_rbd}). Lastly, almost surely%
\begin{eqnarray*}
\left\Vert \mathbf{\Phi }_{b}\right\Vert &=&\left\Vert \mathbf{S}%
_{b}n^{-1}\tsum\nolimits_{i=1}^{n}\left[ b^{\prime }\left\{ m\left( \mathbf{T%
}_{i},\mathbf{X}_{i}\right) \right\} -b^{\prime }\left\{ \bar{m}\left( 
\mathbf{T}_{i},\mathbf{X}_{i}\right) \right\} \right] \mathbf{B}\left( 
\mathbf{T}_{i},\mathbf{X}_{i}\right) \right\Vert \\
&\leq &C_{\mathbf{S,}b}\left\Vert n^{-1}\tsum\nolimits_{i=1}^{n}\left[
b^{\prime }\left\{ m\left( \mathbf{T}_{i},\mathbf{X}_{i}\right) \right\}
-b^{\prime }\left\{ \bar{m}\left( \mathbf{T}_{i},\mathbf{X}_{i}\right)
\right\} \right] \mathbf{B}\left( \mathbf{T}_{i},\mathbf{X}_{i}\right)
\right\Vert =\CO_{a.s.}\left( H^{2}\right)
\end{eqnarray*}%
and%
\begin{eqnarray*}
\left\Vert \mathbf{\Phi }_{v}\right\Vert &=&\left\Vert \mathbf{S}%
_{b}n^{-1}\tsum\nolimits_{i=1}^{n}\left[ \sigma \left( \mathbf{T}_{i},%
\mathbf{X}_{i}\right) \varepsilon _{i}\right] \mathbf{B}\left( \mathbf{T}%
_{i},\mathbf{X}_{i}\right) \right\Vert \\
&\leq &C_{\mathbf{S},b}\left\Vert n^{-1}\tsum\nolimits_{i=1}^{n}\left[
\sigma \left( \mathbf{T}_{i},\mathbf{X}_{i}\right) \varepsilon _{i}\right] 
\mathbf{B}\left( \mathbf{T}_{i},\mathbf{X}_{i}\right) \right\Vert =\CO%
_{a.s.}\left( H^{-1/2}n^{-1/2}\log ^{2}n\right) ,
\end{eqnarray*}%
which completes the proof of the lemma. $\hfill \square $

\bigskip

\noindent \textbf{Proof of Theorem 4 }(i) The Mean Value Theorem implies the
existence of $\mathbf{\breve{\beta}}$ between $\mathbf{\hat{\beta}}$ and $%
\mathbf{\tilde{\beta}}$ such that $\left( \mathbf{\hat{\beta}}-\mathbf{%
\tilde{\beta}}\right) =-\left\{ \nabla ^{2}\hat{\ell}_{\mathbf{\beta }%
}\left( \mathbf{\breve{\beta}}\right) \right\} ^{-1}\nabla \hat{\ell}_{%
\mathbf{\beta }}\left( \mathbf{\tilde{\beta}}\right) ,$where 
\begin{equation*}
-\nabla ^{2}\hat{\ell}_{\mathbf{\beta }}\left( \mathbf{\breve{\beta}}\right)
=n^{-1}\sum_{i=1}^{n}b^{\prime \prime }\left\{ \mathbf{\check{\beta}}^{\top }%
\mathbf{T}_{i}+m\left( \mathbf{X}_{i}\right) \right\} \mathbf{T}_{i}\mathbf{T%
}_{i}^{\top }>c_{b}\mathbf{I}_{d_{1}\times d_{1}}
\end{equation*}%
according to Assumption (A6). We have 
\begin{eqnarray}
\nabla \hat{\ell}_{\mathbf{\beta }}\left( \mathbf{\tilde{\beta}}\right)
&=&\left\{ \frac{\partial \hat{\ell}_{\mathbf{\beta }}\left( \mathbf{\tilde{%
\beta}}\right) }{\partial \beta _{k}}\right\} _{k=0}^{d_{1}}=\nabla \hat{\ell%
}_{\mathbf{\beta }}\left( \mathbf{\tilde{\beta}}\right) -\nabla \tilde{\ell}%
_{\mathbf{\beta }}\left( \mathbf{\tilde{\beta}}\right)  \label{EQ: lbeta} \\
&=&n^{-1}\tsum\nolimits_{i=1}^{n}\left[ b^{\prime }\left\{ \mathbf{\tilde{%
\beta}^{\top }T}_{i}+m\left( \mathbf{X}_{i}\right) \right\} -b^{\prime
}\left\{ \mathbf{\tilde{\beta}^{\top }T}_{i}+\hat{m}\left( \mathbf{X}%
_{i}\right) \right\} \right] \mathbf{T}_{i}.  \notag
\end{eqnarray}%
So for a given $0\leq k\leq d_{1}$,%
\begin{align*}
\frac{\partial \hat{\ell}_{\mathbf{\beta }}\left( \mathbf{\tilde{\beta}}%
\right) }{\partial \beta _{k}}& =n^{-1}\tsum\nolimits_{i=1}^{n}\left[
b^{\prime }\left\{ \mathbf{\tilde{\beta}^{\top }T}_{i}+m\left( \mathbf{X}%
_{i}\right) \right\} -b^{\prime }\left\{ \mathbf{\tilde{\beta}^{\top }T}_{i}+%
\hat{m}\left( \mathbf{X}_{i}\right) \right\} \right] T_{ik} \\
& =n^{-1}\tsum\nolimits_{i=1}^{n}b^{\prime \prime }\left\{ \mathbf{\tilde{%
\beta}^{\top }T}_{i}+m\left( \mathbf{X}_{i}\right) \right\} \left\{ m\left( 
\mathbf{X}_{i}\right) -\hat{m}\left( \mathbf{X}_{i}\right) \right\} T_{ik} \\
& +\CO\left[ n^{-1}\tsum\nolimits_{i=1}^{n}\left\{ m\left( \mathbf{X}%
_{i}\right) -\hat{m}\left( \mathbf{X}_{i}\right) \right\} ^{2}T_{ik}\right]
\\
& =I_{k}+\CO_{a.s.}\left( H^{3}+H^{-2}n^{-1}\log n\right) ,
\end{align*}%
by Lemma \ref{LEM:mhat-m}, where $I_{k}=I_{k1}+I_{k2},$%
\begin{equation*}
I_{k1}=n^{-1}\tsum\nolimits_{i=1}^{n}b^{\prime \prime }\left\{ \mathbf{%
\tilde{\beta}^{\top }T}_{i}+m\left( \mathbf{X}_{i}\right) \right\} \left\{
m\left( \mathbf{X}_{i}\right) -\bar{m}\left( \mathbf{X}_{i}\right) \right\}
T_{ik},
\end{equation*}%
\begin{equation*}
I_{k2}=n^{-1}\tsum\nolimits_{i=1}^{n}b^{\prime \prime }\left\{ \mathbf{%
\tilde{\beta}^{\top }T}_{i}+m\left( \mathbf{X}_{i}\right) \right\} \left\{ 
\bar{m}\left( \mathbf{X}_{i}\right) -\hat{m}\left( \mathbf{X}_{i}\right)
\right\} T_{ik}.
\end{equation*}%
According to Lemma \ref{LEM:splineapprox}, $I_{k1}=\CO_{a.s.}\left(
H^{2}\right) $, while 
\begin{eqnarray*}
I_{k2} &=&n^{-1}\tsum\nolimits_{i=1}^{n}b^{\prime \prime }\left\{ \mathbf{%
\tilde{\beta}^{\top }T}_{i}+m\left( \mathbf{X}_{i}\right) \right\} \left\{
\tsum\nolimits_{1\leq J\leq N+1,1\leq \alpha \leq d_{2}}\left( \hat{\lambda}%
_{J,\alpha }-\bar{\lambda}_{J,\alpha }\right) B_{J,\alpha }\left( X_{i\alpha
}\right) \right\} T_{ik} \\
&=&I_{k2,b}+I_{k2,v}+I_{k2,r}
\end{eqnarray*}%
where%
\begin{eqnarray*}
I_{k2,b} &=&n^{-1}\tsum\nolimits_{i=1}^{n}b^{\prime \prime }\left\{ \mathbf{%
\tilde{\beta}^{\top }T}_{i}+m\left( \mathbf{X}_{i}\right) \right\} \left\{
\tsum\nolimits_{1\leq J\leq N+1,1\leq \alpha \leq d_{2}}\Phi _{b,J,\alpha
}B_{J,\alpha }\left( X_{i\alpha }\right) \right\} T_{ik}, \\
I_{k2,v} &=&n^{-1}\tsum\nolimits_{i=1}^{n}b^{\prime \prime }\left\{ \mathbf{%
\tilde{\beta}^{\top }T}_{i}+m\left( \mathbf{X}_{i}\right) \right\} \left\{
\tsum\nolimits_{1\leq J\leq N+1,1\leq \alpha \leq d_{2}}\Phi _{v,J,\alpha
}B_{J,\alpha }\left( X_{i\alpha }\right) \right\} T_{ik}, \\
I_{k2,r} &=&n^{-1}\tsum\nolimits_{i=1}^{n}b^{\prime \prime }\left\{ \mathbf{%
\tilde{\beta}^{\top }T}_{i}+m\left( \mathbf{X}_{i}\right) \right\} \left\{
\tsum\nolimits_{1\leq J\leq N+1,1\leq \alpha \leq d_{2}}\Phi _{r,J,\alpha
}B_{J,\alpha }\left( X_{i\alpha }\right) \right\} T_{ik}.
\end{eqnarray*}%
We have%
\begin{eqnarray*}
\left\vert I_{k2,b}\right\vert &\leq
&C_{b}n^{-1}\tsum\nolimits_{i=1}^{n}\left\{ \tsum\nolimits_{1\leq J\leq
N+1,1\leq \alpha \leq d_{2}}\left\vert \Phi _{b,J,\alpha }\right\vert
\left\vert B_{J,\alpha }\left( X_{i\alpha }\right) \right\vert \right\}
T_{ik} \\
&\leq &C_{\mathbf{Q}}C_{b}\left\{ \tsum\nolimits_{1\leq J\leq N+1,1\leq
\alpha \leq d_{2}}\Phi _{b,J,\alpha }^{2}\right\} ^{1/2}\times \left[
1+\tsum\nolimits_{1\leq J\leq N+1,1\leq \alpha \leq d_{2}}\left\{
n^{-1}\tsum\nolimits_{i=1}^{n}\left\vert B_{J,\alpha }\left( X_{i\alpha
}\right) \right\vert \right\} ^{2}\right] ^{1/2} \\
&=&C_{\mathbf{Q}}C_{b}\times \CO_{a.s.}\left( N_{d}^{1/2}H^{5/2}\right)
\times \left\{ \CO_{a.s.}\left( N+1\right) \times d_{2}\times \CO%
_{a.s.}\left( H\right) \right\} \\
&=&\CO_{a.s.}\left( H^{2}\right) =\Co_{a.s.}\left( n^{-1/2}\right) .
\end{eqnarray*}%
according to (\ref{EQ:lamdahat-lamdabar}). Similarly, 
\begin{equation*}
\left\vert I_{k2,r}\right\vert =\CO_{p}\left(
N_{d}H^{7/2}+N_{d}H^{-1/2}n^{-1}\log n\right) =\Co_{p}\left( n^{-1/2}\right)
.
\end{equation*}%
We have $I_{k2,v}=\widetilde{I}_{k2,v}+\CO_{a.s.}\left( n^{-1/2}\right)
\times \CO_{a.s.}\left( N_{d}^{1/2}n^{-1/2}\log n\right) \times \CO\left(
N\right) $, where 
\begin{equation*}
\widetilde{I}_{k2,v}=n^{-1}\tsum\nolimits_{i=1}^{n}b^{\prime \prime }\left\{
m\left( \mathbf{T}_{i},\mathbf{X}_{i}\right) \right\} \left\{
\tsum\nolimits_{1\leq J\leq N+1,1\leq \alpha \leq d_{2}}\Phi _{v,J,\alpha
}B_{J,\alpha }\left( X_{i\alpha }\right) \right\} T_{ik}
\end{equation*}%
\begin{equation*}
=-n^{-1}\tsum\nolimits_{i=1}^{n}b^{\prime \prime }\left\{ m\left( \mathbf{T}%
_{i},\mathbf{X}_{i}\right) \right\} n^{-1}\tsum\nolimits_{i^{\prime
}=1}^{n}\sigma \left( \mathbf{T}_{i^{\prime }},\mathbf{X}_{i^{\prime
}}\right) \varepsilon _{i^{\prime }}\mathbf{B}^{\top }\left( \mathbf{X}%
_{i^{\prime }}\right) \mathbf{S}_{b\_\mathbf{\beta }}\mathbf{B}\left( 
\mathbf{X}_{i}\right) T_{ik}
\end{equation*}%
where $\mathbf{B}\left( \mathbf{x}\right) =\left\{ B_{1,1}\left(
x_{1}\right) ,\ldots,B_{N+1,d_{2}}\left( x_{d_{2}}\right) \right\} ^{\top }$
and $\mathbf{S}_{b\_\mathbf{\beta }}$ consists of columns $2+d_{1}$ to $%
N_{d} $ of $\mathbf{S}_{b}$ defined in (\ref{DEF:VbSb}). $\widetilde{I}%
_{k2,v}=\Co_{a.s.}\left( n^{-1/2}\right) $ by calculation similarly to the
proof of Theorem 5 in [11]. Putting the above together, one has%
\begin{equation*}
\left\vert \mathbf{\hat{\beta}}-\mathbf{\tilde{\beta}}\right\vert =\Co%
_{p}\left( n^{-1/2}\right) .
\end{equation*}

\noindent (ii) According to Section 11.2 in [15], 
\begin{equation}
w_{i}=\frac{1}{n}\frac{1}{1+\lambda \left( \mathbf{\beta }\right) ^{\top
}Z_{i}\left( \mathbf{\beta }\right) }\text{,}\frac{1}{n}\tsum%
\nolimits_{i=1}^{n}\frac{Z_{i}\left( \mathbf{\beta }\right) }{1+\lambda
\left( \mathbf{\beta }\right) ^{\top }Z_{i}\left( \mathbf{\beta }\right) }=0,
\label{EQ:elquation}
\end{equation}%
where $Z_{i}\left( \mathbf{\beta }\right) =\left[ Y_{i}-b^{\prime }\left\{ 
\mathbf{\beta }^{\top }\mathbf{T}_{i}+m\left( \mathbf{X}_{i}\right) \right\} %
\right] \mathbf{T}_{i}=$ $\sigma \left( \mathbf{T}_{i},\mathbf{X}_{i}\right)
\varepsilon _{i}\mathbf{T}_{i}.$

$n^{-1}\tsum\nolimits_{i=1}^{n}\sigma \left( \mathbf{T}_{i},\mathbf{X}%
_{i}\right) \varepsilon _{i}\mathbf{T}_{i}\overset{\tciLaplace }{\rightarrow 
}N\left( \mathbf{0},a\left( \phi \right) \left[ \E b^{\prime \prime }\left\{
m\left( \mathbf{T},\mathbf{X}\right) \right\} \mathbf{TT}^{\top }\right]
^{-1}\right) $ by central limit theorem and 
\begin{equation*}
\mathrm{max}_{1\leq i\leq n}\left\vert \lambda \left( \mathbf{\beta }\right)
^{\top }Z_{i}\left( \mathbf{\beta }\right) \right\vert =o_{p}\left( 1\right)
.
\end{equation*}%
So%
\begin{eqnarray*}
-2\log \tilde{R}\left( \mathbf{\beta }\right) &=&-2\Sigma _{i=1}^{n}\log
\left( nw_{i}\right) \\
&=&2\Sigma _{i=1}^{n}\log \left[ 1+\lambda \left( \mathbf{\beta }\right)
^{\top }Z_{i}\left( \mathbf{\beta }\right) \right] \\
&=&2\Sigma _{i=1}^{n}\left\{ \lambda \left( \mathbf{\beta }\right) ^{\top
}Z_{i}\left( \mathbf{\beta }\right) \right\} -\Sigma _{i=1}^{n}\left\{
\lambda \left( \mathbf{\beta }\right) ^{\top }Z_{i}\left( \mathbf{\beta }%
\right) \right\} ^{2}+2\Sigma _{i=1}^{n}\eta _{i}
\end{eqnarray*}%
where $\eta _{i}=\CO_{p}\left( \left\{ \lambda ^{\top }\left( \mathbf{\beta }%
\right) Z_{i}\left( \mathbf{\beta }\right) \right\} ^{3}\right) $ with 
\begin{equation*}
\left\vert \Sigma _{i=1}^{n}\eta _{i}\right\vert \leq C\left\Vert \lambda
\left( \mathbf{\beta }\right) ^{\top }\right\Vert ^{3}\Sigma
_{i=1}^{n}\left\Vert Z_{i}\left( \mathbf{\beta }\right) \right\Vert ^{3}=\Co%
_{p}\left( 1\right) .
\end{equation*}%
Similarly, $-2\log \hat{R}\left( \mathbf{\beta }\right) =2\Sigma
_{i=1}^{n}\left\{ \hat{\lambda}^{\top }\left( \mathbf{\beta }\right) \hat{Z}%
_{i}\left( \mathbf{\beta }\right) \right\} -\Sigma _{i=1}^{n}\left\{ \hat{%
\lambda}\left( \mathbf{\beta }\right) ^{\top }\hat{Z}_{i}\left( \mathbf{%
\beta }\right) \right\} ^{2}$ with%
\begin{equation*}
\hat{Z}_{i}\left( \mathbf{\beta }\right) =\left[ Y_{i}-b^{\prime }\left\{ 
\mathbf{\beta }^{\top }\mathbf{T}_{i}+\hat{m}\left( \mathbf{X}_{i}\right)
\right\} \right] \mathbf{T}_{i}.
\end{equation*}%
So the difference

\begin{eqnarray*}
&&-2\log \hat{R}\left( \mathbf{\beta }\right) +2\log \tilde{R}\left( \mathbf{%
\beta }\right) \\
&=&2\Sigma _{i=1}^{n}\left\{ \hat{\lambda}\left( \mathbf{\beta }\right)
^{\top }\hat{Z}_{i}\left( \mathbf{\beta }\right) \right\} -\Sigma
_{i=1}^{n}\left\{ \hat{\lambda}\left( \mathbf{\beta }\right) ^{\top }\hat{Z}%
_{i}\left( \mathbf{\beta }\right) \right\} ^{2}+ \\
&&-2\Sigma _{i=1}^{n}\left\{ \lambda ^{\top }\left( \mathbf{\beta }\right)
Z_{i}\left( \mathbf{\beta }\right) \right\} +\Sigma _{i=1}^{n}\left\{
\lambda \left( \mathbf{\beta }\right) ^{\top }Z_{i}\left( \mathbf{\beta }%
\right) \right\} ^{2}+\Co_{p}\left( 1\right) \\
&=&2I_{1}+I_{2}+\Co_{p}\left( 1\right)
\end{eqnarray*}%
with%
\begin{eqnarray*}
I_{1} &=&\Sigma _{i=1}^{n}\left\{ \hat{\lambda}\left( \mathbf{\beta }\right)
^{\top }\hat{Z}_{i}\left( \mathbf{\beta }\right) -\lambda \left( \mathbf{%
\beta }\right) ^{\top }Z_{i}\left( \mathbf{\beta }\right) \right\} , \\
I_{2} &=&\Sigma _{i=1}^{n}\left[ \left\{ \lambda \left( \mathbf{\beta }%
\right) ^{\top }Z_{i}\left( \mathbf{\beta }\right) \right\} ^{2}-\left\{ 
\hat{\lambda}\left( \mathbf{\beta }\right) ^{\top }\hat{Z}_{i}\left( \mathbf{%
\beta }\right) \right\} ^{2}\right] .
\end{eqnarray*}%
Rewrite 
\begin{equation*}
I_{1}=\Sigma _{i=1}^{n}\hat{\lambda}\left( \mathbf{\beta }\right) ^{\T%
}\left\{ \hat{Z}_{i}\left( \mathbf{\beta }\right) -Z_{i}\left( \mathbf{\beta 
}\right) \right\} +\Sigma _{i=1}^{n}\left\{ \hat{\lambda}\left( \mathbf{%
\beta }\right) -\lambda \left( \mathbf{\beta }\right) \right\} ^{\T%
}Z_{i}\left( \mathbf{\beta }\right) ,
\end{equation*}%
then%
\begin{eqnarray*}
&&\Sigma _{i=1}^{n}\hat{\lambda}\left( \mathbf{\beta }\right) ^{\top
}\left\{ \hat{Z}_{i}\left( \mathbf{\beta }\right) -Z_{i}\left( \mathbf{\beta 
}\right) \right\} \\
&=&\hat{\lambda}\left( \mathbf{\beta }\right) ^{\top }\Sigma
_{i=1}^{n}\left\{ \hat{Z}_{i}\left( \mathbf{\beta }\right) -Z_{i}\left( 
\mathbf{\beta }\right) \right\} \\
&=&\hat{\lambda}\left( \mathbf{\beta }\right) ^{\top }\Sigma _{i=1}^{n}\left[
b^{\prime }\left\{ \mathbf{\beta }^{\top }\mathbf{T}_{i}+m\left( \mathbf{X}%
_{i}\right) \right\} -b^{\prime }\left\{ \mathbf{\beta }^{\top }\mathbf{T}%
_{i}+\hat{m}\left( \mathbf{X}_{i}\right) \right\} \right] \mathbf{T}_{i} \\
&\leq &\left\Vert \hat{\lambda}\left( \mathbf{\beta }\right) ^{\T%
}\right\Vert \left\Vert \Sigma _{i=1}^{n}\left[ b^{\prime }\left\{ \mathbf{%
\beta }^{\top }\mathbf{T}_{i}+m\left( \mathbf{X}_{i}\right) \right\}
-b^{\prime }\left\{ \mathbf{\beta }^{\top }\mathbf{T}_{i}+\hat{m}\left( 
\mathbf{X}_{i}\right) \right\} \right] \mathbf{T}_{i}\right\Vert \\
&=&\CO_{p}\left( n^{-1/2}\right) \Co_{p}\left( n^{1/2}\right) =\Co_{p}\left(
1\right)
\end{eqnarray*}%
following $\left\Vert \hat{\lambda}\left( \mathbf{\beta }\right) ^{\T%
}\right\Vert =\CO_{p}\left( n^{-1/2}\right) $ and 
\begin{equation*}
\Sigma _{i=1}^{n}\left[ b^{\prime }\left\{ \mathbf{\beta }^{\top }\mathbf{T}%
_{i}+m\left( \mathbf{X}_{i}\right) \right\} -b^{\prime }\left\{ \mathbf{%
\beta }^{\top }\mathbf{T}_{i}+\hat{m}\left( \mathbf{X}_{i}\right) \right\} %
\right] \mathbf{T}_{i}=\Co_{p}\left( n^{1/2}\right)
\end{equation*}%
from the proof for (\ref{EQ: lbeta}). Denote%
\begin{eqnarray*}
\hat{S}\left( \mathbf{\beta }\right) &=&\frac{1}{n}\Sigma _{i=1}^{n}\left[
Y_{i}-b^{\prime }\left\{ \mathbf{\beta }^{\top }\mathbf{T}_{i}+\hat{m}\left( 
\mathbf{X}_{i}\right) \right\} \right] ^{2}\mathbf{T}_{i}\mathbf{T}%
_{i}{}^{\top }, \\
S\left( \mathbf{\beta }\right) &=&\frac{1}{n}\Sigma _{i=1}^{n}\left[
Y_{i}-b^{\prime }\left\{ \mathbf{\beta }^{\top }\mathbf{T}_{i}+m\left( 
\mathbf{X}_{i}\right) \right\} \right] ^{2}\mathbf{T}_{i}\mathbf{T}%
_{i}{}^{\top }.
\end{eqnarray*}%
According to Section 11.2 in [15], 
\begin{eqnarray*}
\hat{\lambda}\left( \mathbf{\beta }\right) &=&\hat{S}^{-1}\left( \mathbf{%
\beta }\right) n^{-1}\Sigma _{i=1}^{n}\hat{Z}_{i}\left( \mathbf{\beta }%
\right) +\Co_{p}\left( n^{-1/2}\right) , \\
\lambda \left( \mathbf{\beta }\right) &=&S^{-1}\left( \mathbf{\beta }\right)
n^{-1}\Sigma _{i=1}^{n}Z_{i}\left( \mathbf{\beta }\right) +\Co_{p}\left(
n^{-1/2}\right) .
\end{eqnarray*}%
We have 
\begin{eqnarray*}
&&\hat{S}\left( \mathbf{\beta }\right) -S\left( \mathbf{\beta }\right) \\
&=&\frac{1}{n}\Sigma _{i=1}^{n}\left\{ \left[ Y_{i}-b^{\prime }\left\{ 
\mathbf{\beta }^{\top }\mathbf{T}_{i}+\hat{m}\left( \mathbf{X}_{i}\right)
\right\} \right] ^{2}-\left[ Y_{i}-b^{\prime }\left\{ \mathbf{\beta }^{\top }%
\mathbf{T}_{i}+m\left( \mathbf{X}_{i}\right) \right\} \right] ^{2}\right\} 
\mathbf{T}_{i}\mathbf{T}_{i}{}^{\top }+\Co_{p}\left( n^{-1/2}\right) \\
&=&\frac{1}{n}\Sigma _{i=1}^{n}\left[ 2Y_{i}-b^{\prime }\left\{ \mathbf{%
\beta }^{\top }\mathbf{T}_{i}+\hat{m}\left( \mathbf{X}_{i}\right) \right\}
-b^{\prime }\left\{ \mathbf{\beta }^{\top }\mathbf{T}_{i}+m\left( \mathbf{X}%
_{i}\right) \right\} \right] \\
&&\left[ b^{\prime }\left\{ \mathbf{\beta }^{\top }\mathbf{T}_{i}+m\left( 
\mathbf{X}_{i}\right) \right\} -b^{\prime }\left\{ \mathbf{\beta }^{\top }%
\mathbf{T}_{i}+\hat{m}\left( \mathbf{X}_{i}\right) \right\} \right] \mathbf{T%
}_{i}\mathbf{T}_{i}{}^{\top }+\Co_{p}\left( n^{-1/2}\right) \\
&=&\frac{1}{n}\Sigma _{i=1}^{n}\left[ 2Y_{i}-2b^{\prime }\left\{ \mathbf{%
\beta }^{\top }\mathbf{T}_{i}+m\left( \mathbf{X}_{i}\right) \right\}
+b^{\prime }\left\{ \mathbf{\beta }^{\top }\mathbf{T}_{i}+m\left( \mathbf{X}%
_{i}\right) \right\} -b^{\prime }\left\{ \mathbf{\beta }^{\top }\mathbf{T}%
_{i}+\hat{m}\left( \mathbf{X}_{i}\right) \right\} \right] \\
&&\left[ b^{\prime }\left\{ \mathbf{\beta }^{\top }\mathbf{T}_{i}+m\left( 
\mathbf{X}_{i}\right) \right\} -b^{\prime }\left\{ \mathbf{\beta }^{\top }%
\mathbf{T}_{i}+\hat{m}\left( \mathbf{X}_{i}\right) \right\} \right] \mathbf{T%
}_{i}\mathbf{T}_{i}{}^{\top }+\Co_{p}\left( n^{-1/2}\right) \\
&=&\frac{1}{n}\Sigma _{i=1}^{n}\left[ 2\sigma \left( \mathbf{T}_{i},\mathbf{X%
}_{i}\right) \varepsilon _{i}+b^{\prime }\left\{ \mathbf{\beta }^{\top }%
\mathbf{T}_{i}+m\left( \mathbf{X}_{i}\right) \right\} -b^{\prime }\left\{ 
\mathbf{\beta }^{\top }\mathbf{T}_{i}+\hat{m}\left( \mathbf{X}_{i}\right)
\right\} \right] \\
&&\left[ b^{\prime }\left\{ \mathbf{\beta }^{\top }\mathbf{T}_{i}+m\left( 
\mathbf{X}_{i}\right) \right\} -b^{\prime }\left\{ \mathbf{\beta }^{\top }%
\mathbf{T}_{i}+\hat{m}\left( \mathbf{X}_{i}\right) \right\} \right] \mathbf{T%
}_{i}\mathbf{T}_{i}{}^{\top }+\Co_{p}\left( n^{-1/2}\right) \\
&=&\frac{1}{n}\Sigma _{i=1}^{n}2\sigma \left( \mathbf{T}_{i},\mathbf{X}%
_{i}\right) \varepsilon _{i}\left[ b^{\prime }\left\{ \mathbf{\beta }^{\top }%
\mathbf{T}_{i}+m\left( \mathbf{X}_{i}\right) \right\} -b^{\prime }\left\{ 
\mathbf{\beta }^{\top }\mathbf{T}_{i}+\hat{m}\left( \mathbf{X}_{i}\right)
\right\} \right] \mathbf{T}_{i}\mathbf{T}_{i}{}^{\top } \\
&&+\frac{1}{n}\Sigma _{i=1}^{n}\left[ b^{\prime }\left\{ \mathbf{\beta }%
^{\top }\mathbf{T}_{i}+m\left( \mathbf{X}_{i}\right) \right\} -b^{\prime
}\left\{ \mathbf{\beta }^{\top }\mathbf{T}_{i}+\hat{m}\left( \mathbf{X}%
_{i}\right) \right\} \right] ^{2}\mathbf{T}_{i}\mathbf{T}_{i}{}^{\top }+\Co%
_{p}\left( n^{-1/2}\right) \\
&=&\Co_{p}\left( n^{-1/2}\right) +\Co_{p}\left( n^{-1/2}\right) ++\Co%
_{p}\left( n^{-1/2}\right) =\Co_{p}\left( n^{-1/2}\right) .
\end{eqnarray*}%
So

\begin{eqnarray*}
&&\hat{\lambda}\left( \mathbf{\beta }\right) -\lambda \left( \mathbf{\beta }%
\right) \\
&=&\hat{S}^{-1}\left( \mathbf{\beta }\right) n^{-1}\Sigma _{i=1}^{n}\hat{Z}%
_{i}\left( \mathbf{\beta }\right) -S^{-1}\left( \mathbf{\beta }\right)
n^{-1}\Sigma _{i=1}^{n}Z_{i}\left( \mathbf{\beta }\right) \\
&=&\hat{S}^{-1}\left( \mathbf{\beta }\right) n^{-1}\Sigma _{i=1}^{n}\left\{ 
\hat{Z}_{i}\left( \mathbf{\beta }\right) -Z_{i}\left( \mathbf{\beta }\right)
\right\} +\left\{ \hat{S}^{-1}\left( \mathbf{\beta }\right) -S^{-1}\left( 
\mathbf{\beta }\right) \right\} n^{-1}\Sigma _{i=1}^{n}Z_{i}\left( \mathbf{%
\beta }\right) \\
&=&\Co_{p}\left( n^{-1/2}\right) .
\end{eqnarray*}%
Then%
\begin{eqnarray*}
&&\Sigma _{i=1}^{n}\left\{ \hat{\lambda}\left( \mathbf{\beta }\right)
-\lambda \left( \mathbf{\beta }\right) \right\} ^{\top}Z_{i}\left( \mathbf{%
\beta }\right) \\
&=&\left\{ \hat{\lambda}\left( \mathbf{\beta }\right) -\lambda \left( 
\mathbf{\beta }\right) \right\} ^{\top}\Sigma _{i=1}^{n}Z_{i}\left( \mathbf{%
\beta }\right) \\
&=&\Co_{p}\left( n^{-1/2}\right) \CO_{p}\left( n^{1/2}\right) =\Co_{p}\left(
1\right) ,
\end{eqnarray*}%
and $I_{1}=\Co_{p}\left( 1\right) $.%
\begin{eqnarray*}
I_{2} &=&\Sigma _{i=1}^{n}\left[ \left\{ \lambda \left( \mathbf{\beta }%
\right) ^{\top}Z_{i}\left( \mathbf{\beta }\right) \right\} ^{2}-\left\{ \hat{%
\lambda}\left( \mathbf{\beta }\right) ^{\top}\hat{Z}_{i}\left( \mathbf{\beta 
}\right) \right\} ^{2}\right] \\
&=&\Sigma _{i=1}^{n}\left\{ \lambda \left( \mathbf{\beta }\right) ^{\T%
}Z_{i}\left( \mathbf{\beta }\right) +\hat{\lambda}\left( \mathbf{\beta }%
\right) ^{\top}\hat{Z}_{i}\left( \mathbf{\beta }\right) \right\} \left\{
\lambda \left( \mathbf{\beta }\right) ^{\top}Z_{i}\left( \mathbf{\beta }%
\right) -\hat{\lambda}\left( \mathbf{\beta }\right) ^{\top}\hat{Z}_{i}\left( 
\mathbf{\beta }\right) \right\} \\
&=&\Co_{p}\left( 1\right) .
\end{eqnarray*}%
by similar proof. Putting together, we have 
\begin{equation*}
-2\log \hat{R}\left( \mathbf{\beta }\right) +2\log \tilde{R}\left( \mathbf{%
\beta }\right) =\Co_{p}\left( 1\right) .
\end{equation*}
$\hfill \square $

\end{document}